\documentclass[conference]{IEEEtran}

\usepackage{amsmath,amsfonts,bm}

\def\eqref#1{equation~\ref{#1}}

\def\1{\bm{1}}

\DeclareMathAlphabet{\mathsfit}{\encodingdefault}{\sfdefault}{m}{sl}
\SetMathAlphabet{\mathsfit}{bold}{\encodingdefault}{\sfdefault}{bx}{n}

\usepackage{hyperref}

\usepackage[utf8]{inputenc} 
\usepackage[T1]{fontenc}    
\usepackage{hyperref}       
\usepackage{url}            
\usepackage{booktabs}       
\usepackage{amsfonts}       
\usepackage{nicefrac}       
\usepackage{microtype}      
\usepackage{xcolor}         

\usepackage{makecell}
\usepackage{booktabs}       
\usepackage{amsfonts}       
\usepackage{amsmath}
\usepackage{amsthm}
\usepackage{amssymb}
\usepackage[capitalize,noabbrev]{cleveref} 
\usepackage{graphicx}
\usepackage{subcaption}
\usepackage{wrapfig} 
\usepackage{multirow}

\usepackage{circledsteps}
\usepackage{pifont}
\usepackage{tikz}

\usepackage{xcolor}         
\usepackage{pifont}
\definecolor{redx}{RGB}{180,0,0}
\definecolor{greenx}{RGB}{0,180,0}

\newcommand{\redxmark}{\color{redx}\ding{55}}
\newcommand{\greencmark}{\color{greenx}\ding{51}}
\definecolor{redx}{RGB}{180,0,0}
\definecolor{greenx}{RGB}{0,180,0}
\usepackage{threeparttable}

\theoremstyle{plain}
\newtheorem{theorem}{Theorem}[section]

\theoremstyle{definition}
\newtheorem{definition}[theorem]{Definition}

\theoremstyle{remark}

\usepackage{algorithm}
\usepackage{algpseudocodex}

\usepackage{xspace}  

 \usepackage{balance}

\newcommand{\name}{{CustodianFL}\xspace}

\usepackage{bbding}

\definecolor{cluci}{RGB}{219, 109, 0}
\definecolor{cltamu}{RGB}{80, 0, 0}
\definecolor{clbham}{RGB}{0, 81, 158}
\definecolor{clusc}{RGB}{0, 130, 80}
\definecolor{clpan}{RGB}{153, 27, 30}
\definecolor{clcmu}{RGB}{85, 43, 111}
\definecolor{LimeGreen}{RGB}{51, 205, 51}

\newcommand{\mkuci}[0]{{\color{cluci}{$^\star$}}}
\newcommand{\mktamu}[0]{{\color{cltamu}{$^\dag$}}}
\newcommand{\mkbham}[0]{{\color{clbham}{$^\ddag$}}}
\newcommand{\mkusc}[0]{{\color{clusc}{$^\S$}}}
\newcommand{\mkpan}[0]{{\color{clpan}{$^\P$}}}
\newcommand{\mkcmu}[0]{{\color{clcmu}{$^\|$}}}

\newcommand{\mkletter}[0]{{\color{LimeGreen}{\Envelope}}}

\newcommand{\colorhref}[3][black]{\href{#2}{\color{#1}{#3}}}%

\renewcommand\footnotemark{}

\begin{document}
\title{Kick Bad Guys Out! Conditionally Activated Anomaly Detection in Federated Learning with Zero-Knowledge Proof Verification}

\author{
    \colorhref{mailto:shanshan.han@uci.edu}{\rm Shanshan Han}\mkuci \mkletter \rm ,
    \colorhref{mailto:ww6726@tamu.edu}{\rm Wenxuan Wu}\mktamu \rm ,
    \colorhref{mailto:b.buyukates@bham.ac.uk}{\rm Baturalp Buyukates}\mkbham \rm ,
    \colorhref{mailto:weizhaoj@usc.edu}{\rm Weizhao Jin}\mkusc \rm , \\
    \colorhref{https://qifanz.com/}{\rm Qifan Zhang}\mkpan \mkletter \rm ,
    \colorhref{mailto:yuhangya@andrew.cmu.edu}{\rm Yuhang Yao}\mkcmu \rm , and
    \colorhref{mailto:avestime@usc.edu}{\rm Salman Avestimehr}\mkusc \rm
    \thanks{This paper appeared in the
    Workshop on Attack Provenance, Reasoning, and Investigation for Security
    in the Monitored Environment (PRISM), co-located with NDSS Symposium 2026.}
    \medskip
    \\
    \mkuci \colorhref{https://uci.edu/}{University of California, Irvine},
    \mktamu \colorhref{https://www.tamu.edu/}{Texas A\&M University},
    \mkbham \colorhref{https://www.birmingham.ac.uk/}{University of Birmingham}
    \\
    \mkusc \colorhref{https://www.usc.edu/}{University of Southern California},
    \mkpan \colorhref{https://www.paloaltonetworks.com/}{Palo Alto Networks},
    \mkcmu \colorhref{https://www.cmu.edu/}{Carnegie Mellon University}
}

\IEEEoverridecommandlockouts
\makeatletter\def\@IEEEpubidpullup{0.5\baselineskip}\makeatother
\IEEEpubid{\parbox{\columnwidth}{
}
\hspace{\columnsep}\makebox[\columnwidth]{}}

\maketitle

\begin{abstract}

Federated Learning (FL) systems are susceptible to adversarial attacks, such as model poisoning attacks and backdoor attacks. Existing defense mechanisms face critical limitations in deployments, such as relying on impractical assumptions (\textit{e}.\textit{g}., adversaries acknowledging the presence of attacks before attacking) or undermining accuracy in model training, even in benign scenarios.
To address these challenges, we propose \name, a two-staged anomaly detection method specifically designed for FL deployments. 
In the first stage, it flags suspicious client activities. In the second stage that is activated only when needed, it further examines these candidates using the $3\sigma$ rule to identify and exclude truly malicious local models from FL training.
To ensure integrity and transparency within the FL system, \name integrates zero-knowledge proofs, enabling clients to cryptographically verify the server's detection process without relying on the server's goodwill. 
\name operates without unrealistic assumptions and avoids interfering with FL training in attack-free scenarios. 
It bridges the gap between theoretical advances in FL security and the practical demands of real FL systems. 
Experimental results demonstrate that {\name consistently delivers performance comparable to benign cases}, highlighting its effectiveness in identifying and eliminating malicious models with high accuracy.

\end{abstract}

\IEEEpeerreviewmaketitle

\section{Introduction}\label{sec: intro}

Federated Learning (FL)~\cite{mcmahan2017communication} is vulnerable to various security threats~\cite{cao2022mpaf,bhagoji2019analyzing,lam2021gradient,jin2021cafe,tomsett2019model,chen2017distributed,label_flipping,Kariyappa2022CocktailPA,Zhang2022NeurotoxinDB}. 
Malicious clients may deliberately manipulate their local models to disrupt global model convergence~\cite{fang2020local,chen2017distributed} or implant backdoors that cause the global model to misclassify specific inputs~\cite{how_to_backdoor,wang2020attack}. 
These threats undermine the reliability of FL systems, as the participation of adversarial participants may be unpredictable  and their malicious intent remains hidden until an attack is successfully executed.

Existing FL defense mechanisms face critical limitations in practical deployment~\cite{krum,cao2022flcert,foolsgold,he2022byzantine,karimireddy2020byzantine,kumari2023baybfed,rfa,sun2019can,fu2019attack,ozdayi2021defending,sun2021fl,yin2018byzantine,chen2017distributed,xie2020slsgd,li2020learning,cao2020fltrust,yu2023g,zhang2024nspfl,nguyen2022flame,rieger2022deepsight,yang2019byzantine}. 
These defenses might rely on impractical assumptions or require unrealistic prior knowledge~\cite{krum,sun2019can,fu2019attack,ozdayi2021defending}. 
Some defenses assume that the server is aware of the number of malicious clients and the timing of attacking~\cite{krum,rfa,foolsgold}, an assumption that rarely holds in practice. 
However, adversaries might conceal their malicious intent and not notify the FL system before attacking. 
Some defenses modify local models and/or model aggregation, \textit{e}.\textit{g}., adjusting aggregation functions~\cite{rfa}, re-weighting client updates or local models~\cite{foolsgold,nguyen2022flame,rieger2022deepsight}, or discarding suspicious models~\cite{krum}, to improve robustness. 
However, these methods degrade model performance even in attack-free settings. By interfering with aggregation aggressively, they penalize honest participants and undermine accuracy across training rounds. This issue is particularly problematic in real-world FL systems, where attacks are infrequent.
Furthermore, these defenses operate on the FL server without providing mechanisms for clients to verify their correct execution~\cite{foolsgold,nguyen2022flame,rieger2022deepsight,krum}. The honest clients have to trust the server blindly, undermining transparency and accountability in FL systems.

To ensure practicality in FL systems, FL defenses must satisfy three requirements:
\textit{{i}}) 
the defense should operate on demand, activated only when attacks might have happened to avoid interference during benign training rounds;
\textit{{ii}}) 
upon detecting a potential attack, the defense should accurately identify malicious local models and mitigate or eliminate their negative impact without harming benign ones; and
\textit{{iii}}) 
the defense should include a verification mechanism that enables clients to validate the integrity of server-side operations without relying solely on the server’s goodwill.

\begin{figure*}
    \centering
    \includegraphics[width=0.92\linewidth]{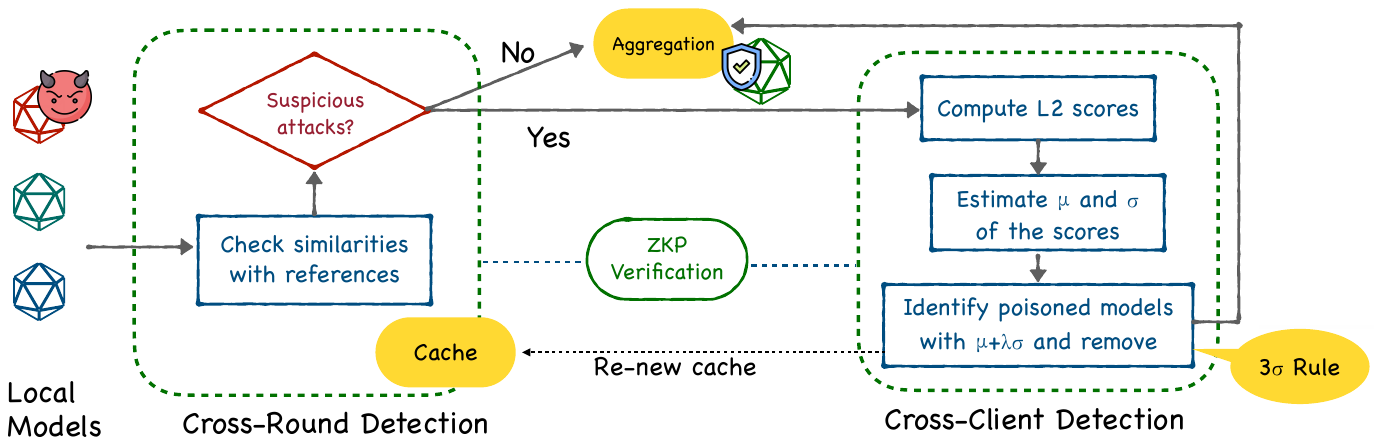}
    \caption{Overview of \name}
    \label{fig:overview}
\end{figure*}

\begin{table*}[ht]
\centering
\caption{Comparison with state-of-the-art methods}
\label{tb:state-of-the-art-compare}
\centering
\small
\resizebox{\textwidth}{!}{%
\begin{tabular}{l|c|c|c|c|c|c|c|c}
\toprule
\textbf{Method} & \makecell[c]{\textbf{Attack} \\ \textbf{presence} \\ \textbf{detection}} & \makecell[c]{\textbf{Removing} \\ \textbf{malicious} \\ \textbf{models}} & \makecell[c]{\textbf{Free from} \\ \textbf{impractical} \\ \textbf{knowledge}} & \makecell[c]{\textbf{Free from} \\ \textbf{reweighting}} & \makecell[c]{\textbf{Free from} \\ \textbf{aggregation} \\ \textbf{modification}} & \makecell[c]{\textbf{Free from} \\ \textbf{harming} \\ \textbf{benign models}} & \makecell[c]{\textbf{\emph{Robust}} \\ \textbf{performance in} \\ \textbf{non-attack scenarios}} & \makecell[c]{\textbf{Execution} \\ \textbf{Integrity} \\ \textbf{Verification}} \\
\midrule
\textbf{Krum}~\cite{krum} & \redxmark & \greencmark & \redxmark & \greencmark & \greencmark & \redxmark & \redxmark & \redxmark \\\midrule
\textbf{RFA}~\cite{rfa} & \redxmark & \redxmark & \greencmark & \greencmark & \redxmark & \redxmark & \redxmark & \redxmark \\\midrule
\textbf{Foolsgold}~\cite{foolsgold} & \redxmark & \redxmark & \greencmark & \redxmark & \greencmark & \redxmark & \redxmark & \redxmark \\\midrule
\textbf{NormClip}~\cite{sun2019can} & \redxmark & \redxmark & \greencmark & \greencmark & \greencmark & \redxmark & \redxmark & \redxmark \\\midrule
\textbf{Bucketing}~\cite{karimireddy2020byzantine} & \redxmark & \redxmark & \greencmark & \redxmark & \greencmark & \redxmark & \redxmark & \redxmark \\\midrule
\textbf{Median}~\cite{yin2018byzantine} & \redxmark & \redxmark & \greencmark & \greencmark & \redxmark & \greencmark & \redxmark & \redxmark \\\midrule
\textbf{TrimMean}~\cite{yin2018byzantine}  & \redxmark & \redxmark & \greencmark & \greencmark & \greencmark & \redxmark & \redxmark & \redxmark \\\midrule
\textbf{Flip}~\cite{zhang2022flip} & \redxmark &\redxmark & \redxmark & \greencmark& \greencmark & \redxmark&\redxmark &\redxmark \\\midrule
\textbf{Snowball}~\cite{snowball} & \redxmark & \greencmark & \greencmark  & \greencmark &\redxmark  & \greencmark&\redxmark &\redxmark \\\midrule

\textbf{Flame}~\cite{nguyen2022flame} & \redxmark & \redxmark & \greencmark  & \greencmark &\redxmark  & \redxmark &\redxmark &\redxmark \\\midrule
\textbf{DeepSight}~\cite{rieger2022deepsight} & \greencmark & \greencmark & \greencmark  & \greencmark &\redxmark  & \redxmark &\redxmark &\redxmark \\\midrule
\textbf{BayBFed}~\cite{kumari2023baybfed} & \redxmark & \greencmark & \greencmark  & \greencmark &\greencmark  & \greencmark &\greencmark &\redxmark \\\midrule


\textbf{\name (Ours)} & \greencmark & \greencmark & \greencmark & \greencmark & \greencmark & \greencmark & \greencmark & \greencmark \\

\bottomrule
\end{tabular}}
\end{table*}


This paper presents \name, a two-stage defense mechanism that detects and filters out malicious client models in each FL training round while addressing the challenges in real-world FL systems. 
As illustrated in~\cref{fig:overview}, \name begins with a \textit{\textbf{cross-round detection}} that monitors round-to-round behaviors to identify suspicious activities of local clients. 
Upon detection of suspicious activities, \name activates \textit{\textbf{cross-client detection}} that quantifies the maliciousness level of each local model and filters out the malicious ones based on the 3$\sigma$ rule~\cite{three_sigma}. 
To ensure transparency and integrity in training, \name integrates Zero-Knowledge Proofs (ZKPs)~\cite{Goldwasser19} to provide cryptographic guarantees of its benign execution on the FL server. 
\cref{tb:state-of-the-art-compare} compares \name against the state-of-the-art methods~\cite{krum,foolsgold,karimireddy2020byzantine,rfa,sun2019can,yin2018byzantine,zhang2022flip,snowball}. 
Our key contributions are as follows:

\noindent\textbf{\textit{i}) Practical Applicability in FL Systems:}
     \name operates without requiring impractical prior knowledge (\textit{e}.\textit{g}., the number of malicious clients or the timing of attacks) and is designed for real-world FL systems. To the best of our knowledge, it is the first method that bridges the gap between academic research and practical applications of FL security.

\noindent{\textbf{\textit{ii}) Conditional Activation:}} 
\name does not interfere with training in attack-free scenarios, a critical requirement for FL deployments where adversarial behavior is infrequent and model accuracy is paramount. It activates detection and filtering only when attacks are suspected, avoiding disturbing benign training and preventing accuracy degradation.

\noindent\textbf{\textit{iii}) Non-Disruptive Operation:}
Upon identification of suspicious activities, \name detects and removes malicious local models with high accuracy, without modifying the aggregation function or impacting benign local models.

\noindent{\textbf{\textit{iv}) Enhanced Detection Accuracy:}}  \name avoids removing local models based on scores directly computed with local models. Instead, it  leverages the statistical properties of the local models and applies the $3\sigma$ rule to identify the malicious ones, thereby enhancing detection accuracy.
    
\noindent\textbf{\textit{v}) Verifiability:} \name enables clients to verify the integrity of server-side operations independently with ZKP, fostering accountability without trusting the FL server blindly.

\section{Problem Setting}

\begin{algorithm}[ht]
    \caption{Krum and $m$-Krum}
    \label{alg:krum}
    \textbf{Input:} $\mathcal{W}$: client submissions of a training round; $m$: number of neighbors considered for computing the Krum score ($m=1$ for standard Krum); $f$: number of malicious clients in each round. \\
    \textbf{Output:} Aggregated global model

    \begin{algorithmic}[1]

    \State $\mathcal{S}_k \gets [\,]$ \Comment{List of Krum scores}
    
    \ForAll{$\mathbf{w}_i \in \mathcal{W}$}
        \State $\mathcal{S}_k(\mathbf{w}_i) \gets \text{compute\_krum\_score}{(\mathcal{W}, i, m, f)}$
    \EndFor
    
    \State $\mathcal{W} \gets \Call{filter}{\mathcal{W}, \mathcal{S}_k}$ 
    \Comment{Keep $|\mathcal{W}|/2$ models with lowest scores}
    \State \Return $\text{average}{(\mathcal{W})}$

    \Function{compute\_krum\_score}{$\mathcal{W}, i, m, f$}
        \State $d \gets [\,]$ \Comment{List of squared distances}
        \State $L \gets |\mathcal{W}|$ \Comment{Total number of clients}

        \ForAll{$\mathbf{w}_j \in \mathcal{W}$}
        \If{$i \neq j$}~$d.\text{append}\left( \| \mathbf{w}_i - \mathbf{w}_j \|^2 \right)$
            \EndIf
        \EndFor

        \State $\text{Sort}(d)$ \Comment{Ascending order}
        \State $\mathcal{S}_k(\mathbf{w}_i) \gets \sum_{k=0}^{L - f - 3} d[k]$
        \Comment{Use the smallest $L - f - 2$ distances}
        \State \Return $\mathcal{S}_k(\mathbf{w}_i)$
    \EndFunction

    \end{algorithmic}
\end{algorithm}

\subsection{Adversary Model}\label{sec:adversarial_setting}

We consider an FL system in which a subset of participating clients may be adversarial and attempt to compromise the training process to achieve some {malicious goals}. Adversarial clients might: 
\textit{i}) inject backdoors into local updates to cause the global model to misclassify specific inputs~\cite{how_to_backdoor,wang2020attack,yu2023g}; 
\textit{ii}) perform Byzantine attacks by intentionally manipulating local models to prevent the global model from converging~\cite{chen2017distributed,fang2020local}; and \textit{iii}) submit fabricated models to the server~\cite{free_rider}. 
We further assume that the adversaries might be \emph{adaptive}, \textit{i}.\textit{e}., they can observe the defense mechanism and adapt their attack strategies accordingly~\cite{wu2023learning}.
We follow common threat model assumptions~\cite{krum,ozdayi2021defending,sun2021fl,yin2018byzantine,chen2017distributed,xie2020slsgd} and assume at least 50\% clients are benign. We assume the FL server is not fully trusted, consistent with deployment scenarios with potentially untrusted execution environments. While clients expect a defense mechanism deployed at the server, they remain uncertain about whether it is executed faithfully.

\subsection{Preliminaries}

\noindent\textbf{Federated Learning (FL). }
FL~\cite{mcmahan2017communication} enables training machine learning models  across decentralized devices using their local data. 
Instead of centralizing data on a single server, FL brings the model to the data, allowing training to be performed locally on each client. 
FL is particularly beneficial when dealing with sensitive data, as the raw data never leaves the local devices during the training process.

\noindent\textbf{Krum.} 
Krum and $m$-Krum select $m$ ($m$ is one in Krum) local models that deviate less from the majority based on their pairwise distances for aggregation, as such local models are more likely to be benign. 
Given that there are $f$ byzantine clients among $L$ clients that participate in each FL iteration, Krum selects one model that is the most likely to be benign as the global model. 
To do so, Krum computes a score for each model $\mathbf{w}_i$, denoted as $\mathcal{S}_K(\mathbf{w}_i)$, using $L-f-2$ local models that are ``closest'' to $\mathbf{w}_i$, and selects the local model with the minimum score to represent the aggregation result.  
For each local model $\mathbf{w}_i$, suppose $C_i^{\mathcal{N}}$ is the set of the $L-f-2$ local models that are closest to $\mathbf{w}_i$,
then $\mathcal{S}_K(\mathbf{w}_i)$ is computed by 
$\mathcal{S}_K(\mathbf{w}_i)=\displaystyle\sum_{j \in \mathcal{C}_i} ||\mathbf{w}_i-\mathbf{w}_j||^2.$ 
The optimization, $m$-Krum~\cite{krum}, selects $m$ local models ($m> 1$) in aggregation. 
The algorithm for Krum and $m$-Krum is summarized in Algorithm~\ref{alg:krum}.

\begin{figure}
    \centering
    \includegraphics[width=0.8\linewidth]{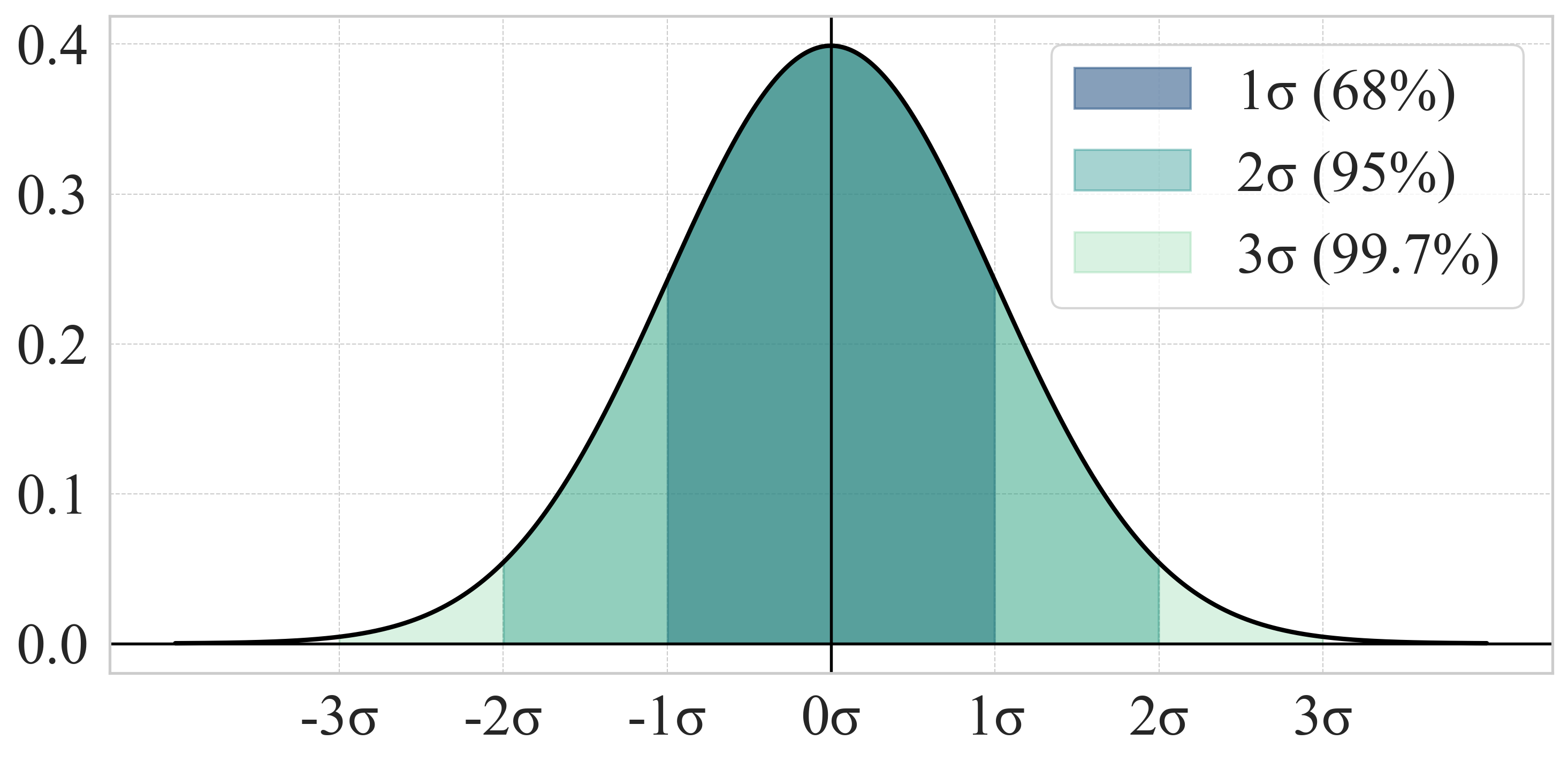}
    \caption{Illustration of the $3\sigma$ rule}
    \label{fig:3sigma}
\end{figure}

\noindent\textbf{3$\sigma$ Rule.} The 3$\sigma$ rule~\cite{three_sigma} is an empirical rule that is commonly used in anomaly detection~\cite{han2019iterative}. It states that approximately 68\%, 95\%, and 99.7\% of data values lie within one, two, and three standard deviations from the mean, respectively, under a normal distribution; see \cref{fig:3sigma}. This rule is broadly applicable in practical settings, as many data distributions approximate normality~\cite{lyon2014normal}. Even when the data is not normally distributed, we can apply transformation to approximate the data to normal distributions~\cite{aoki1950stability, osborne2010improving, sakia1992box, weisberg2001yeo}.

\noindent\textbf{Zero-Knowledge Proofs (ZKPs).} A ZKP~\cite{Goldwasser19} is a proof system enabling a prover to convince a verifier that a function has been correctly computed on the prover's secret input (witness). ZKPs have three properties: \textit{i}) \textit{correctness}: the proof they produce should pass verification if the prover is honest; \textit{ii}) \textit{soundness}: a cheating prover cannot convince the verifier with overwhelming probability, and \textit{iii}) \textit{zero-knowledge}: the prover's witness is not learned by the verifier.

\section{\name: Two-Staged Anomaly Detection}

\name operates in each FL round after the server has collected the local models. It first performs a lightweight \textit{cross-round detection} to assess the likelihood of potential attacks. If suspicious activity is detected, \name then activates a more rigorous \textit{cross-client detection} to evaluate the maliciousness, \textit{i}.\textit{e}., the \textit{evilness level}, of each local model. Models identified as malicious are subsequently removed using the $3\sigma$ rule to mitigate their impact on the global model.

\subsection{Cross-Round Detection}\label{sec:cross_round_check}

Cross-round detection serves as a ``gatekeeper'' and evaluates the likelihood of suspicious activities in local models, such that \name can decide whether to activate the next phase for more rigorous detection or not.

Cross-round detection 
computes cosine similarities between the local models of the current round and some reference models.
Two types of reference models are used, including
\textit{i}) the global model from the previous FL training round; and \textit{ii}) verified benign local models identified in earlier rounds.
These reference models have a high likelihood of being benign thus can serve as a reliable \textit{golden truth} for the cross-round check.
The global model provides a reference for convergence; local models that deviate significantly from the expected global model might be attempting to disrupt training. Meanwhile, comparing clients' current submissions with their previously verified benign submissions enables the detection of behavioral shifts in the current round, \textit{e}.\textit{g}., transitioning from benign to malicious behaviors, thereby flagging inconsistencies in client-specific activities across consecutive FL rounds.

\begin{figure}[h] 
    \centering
    \includegraphics[width=\linewidth]{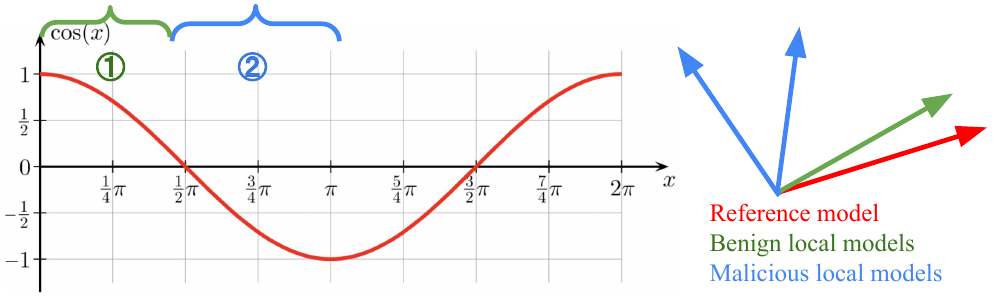}
    
    \caption{Cosine similarities.  \textcolor{green}{\Circled{1}} indicates likely benign models with high cosine similarity, and \textcolor{blue}{\Circled{2}} indicates likely malicious models with low cosine similarity.}
    \label{fig:cosine}
\end{figure}

We illustrate the  idea in~\cref{fig:cosine}. Benign local models are expected to exhibit high similarities to the reference models. 
For each local model $\mathbf{w}_i$ and a reference model $\mathbf{w}_r$, we compute the cosine similarity as
$\mathcal{S}_c(\mathbf{w}_i, \mathbf{w}_r) = \frac{\mathbf{w}_i\cdot\mathbf{w}_r}{||\mathbf{w}_i||\cdot||\mathbf{w}_r||}$.
A high similarity reflects strong alignment between $\mathbf{w}_i$ and $\mathbf{w}_r$, indicating the client is more likely to be benign. Lower similarities, on the other hand, signal potential adversarial behaviors, as malicious clients might submit manipulated models that diverge $\mathbf{w}_i$ from $\mathbf{w}_r$~\cite{chen2017distributed,fang2020local,how_to_backdoor,wang2020attack}. In practice, \name employs a threshold $\gamma$ ($-1<\gamma<1$). Similarity scores lower than $\gamma$ indicate potential adversarial behaviors of the corresponding clients and will activate a further inspection in the second phase, as described later in \S\ref{sec: cross_client}.

The cross-round detection algorithm is summarized in Algorithm~\ref{algo:cross_round}. 
Upon the server receiving local models from clients, it first 
loads the reference models, including the  
client models from the earlier rounds and the global model from the last round. 
Then, it computes cosine similarities between each local model and the corresponding reference. 
Local models exhibiting higher similarities to these references are more likely to be benign, while those with lower similarities are considered suspicious, requiring a rigorous cross-client detection in the next phase.
We note that \name just flags suspicious models without removing them, thus, \name does not rely heavily on cosine similarities.

We leverage a modified Positive Predictive Value (\textsf{PPV})~\cite{fletcher2019clinical} to evaluate the accuracy of  cross-round detection while revealing whether all malicious models are identified.









\begin{algorithm}[ht]
    \caption{\name -Phase 1: Cross-Round Detection}
    \label{algo:cross_round}
    \textbf{Input:} $\tau$: training round ID ($\tau = 0, 1, 2, \ldots$); $\mathcal{W}^{\tau}$: client models of round $\tau$; $\gamma$: similarity threshold. 

    \begin{algorithmic}[1]
	\If{$\tau = 0$}~\Return \textbf{True} \Comment{No previous round, activate cross-client detection by default}
        \EndIf


        \State $\mathcal{W}^{\tau - 1} \gets \text{get\_cached\_client\_models}()$, $\mathbf{w}_g^{\mathit{ref}} \gets \text{get\_global\_model\_of\_last\_round}()$

        \ForAll{$\mathbf{w}_i^{\tau} \in \mathcal{W}^{\tau}$} 
            \State $\mathcal{S}_c(\mathbf{w}_i^{\tau - 1}, \mathbf{w}_i^{\tau}) \gets \text{get\_similarity}(\mathbf{w}_i^{\tau - 1}, \mathbf{w}_i^{\tau})$, $\mathcal{S}_c(\mathbf{w}_g^{\mathit{ref}}, \mathbf{w}_i^{\tau}) \gets \text{get\_similarity}(\mathbf{w}_g^{\mathit{ref}}, \mathbf{w}_i^{\tau})$
		\If{$\mathcal{S}_c(\mathbf{w}_g^{\mathit{ref}}, \mathbf{w}_i^{\tau}) < \gamma$ \textbf{ or } $\mathcal{S}_c(\mathbf{w}_i^{\tau - 1}, \mathbf{w}_i^{\tau}) < \gamma$}~\Return \textbf{True} \Comment{Potential attacks}
            \EndIf
        \EndFor

        \State \Return \textbf{False} \Comment{No suspicious attacks}
    \end{algorithmic}
\end{algorithm}

\begin{definition}
Let $\mathcal{T}$ be an FL training process consisting of $\tau$ rounds ($\tau > 0$). Denote by $\mathcal{W}$ the set of all client submissions across all rounds, partitioned into malicious submissions $\mathcal{W}_{\mathsf{bad}}$ and benign submissions $\mathcal{W}_{\mathsf{good}}$. Let $\mathcal{W}_{\mathsf{bad}}^d$ and $\mathcal{W}_{\mathsf{good}}^d$ represent the sets of submissions detected as \textit{malicious} and \textit{benign}, respectively, by a detection mechanism $\mathcal{M}$. Then true-positives (\textsf{TP}) is defined as $N_\mathsf{TP} = \left| \mathcal{W}_\mathsf{bad}^d \cap \mathcal{W}_\mathsf{bad} \right|$, and 
false-positives (\textsf{FP}) is defined as
$\mathsf{N}_\mathsf{FP}=\left|\mathcal{W}_\mathsf{bad}^d\cap \mathcal{W}_\mathsf{good}\right|$. We define a modified \textsf{PPV} as
$\textsf{PPV}= \frac{\mathsf{N}_{\mathsf{TP}}}{\mathsf{N}_{\mathsf{TP}} + \mathsf{N}_{\mathsf{FP}} + \left|\mathcal{W}_{\mathsf{bad}}\right|}$, where $0\leq \textsf{PPV} \leq \frac{1}{2}.$
\end{definition}

\noindent Ideally, $\textsf{PPV}$ is $\frac{1}{2}$, indicating perfect performance, \textit{i}.\textit{e}., all malicious models are identified ($\mathsf{N}_{\mathsf{TP}} = \left| \mathcal{W}_{\mathsf{bad}} \right|$) and no benign models are misclassified ($\mathsf{N}_{\mathsf{FP}} = 0$).

\begin{proof}
We leverage a modified \textsf{PPV} evaluate the accuracy of the cross-round detection in identifying potential attacks across all FL 
training rounds.
Below, we show that the upper bound of the modified \textsf{PPV} is $\frac{1}{2}$. We have 
    $\textsf{PPV} = \frac{\mathsf{N}_{\mathsf{TP}}}{\mathsf{N}_{\mathsf{TP}} + \mathsf{N}_{\mathsf{FP}} + \left|\mathcal{W}_\mathsf{bad}\right|}$, thus we have $\frac{1}{\textsf{PPV}}=1 + \frac{\mathsf{N}_{\mathsf{FP}}}{\mathsf{N}_{\mathsf{TP}}} + \frac{\left|\mathcal{W}_\mathsf{bad}\right|}{\mathsf{N}_{\mathsf{TP}}}$. As $\frac{\mathsf{N}_{\mathsf{FP}}}{\mathsf{N}_{\mathsf{TP}}} \geq 0$ and $\frac{\left|\mathcal{W}_\mathsf{bad}\right|}{\mathsf{N}_{\mathsf{TP}}}\geq 1$, we have $\frac{1}{\textsf{PPV}}\geq 2$, thus $\textsf{PPV}\leq \frac{1}{2}$.
\end{proof}

\subsection{Cross-Client Detection }\label{sec: cross_client}
Cross-client detection is activated when the cross-round detection flags potential threats, aiming to further verify whether actual attacks have occurred in the current FL round. For each local model, it evaluates its \textit{evilness level} with an $\mathsf{L_2}$ score to measure its deviation. It then applies the $3\sigma$ rule to these scores to identify outliers. These outliers are treated as malicious models and are excluded from aggregation.

The cross-client detection is described in Algorithm~\ref{alg:cross_client}. For each local model $\mathbf{w}_i^\tau$ in the current round $\tau$, the algorithm computes its \textit{evilness level} with an $\mathsf{L_2}$ score as 
$||\mathbf{w}_i^\tau - \mathbf{w}_{\text{g}}^{\tau-1}||_2$, where $\mathbf{w}_{\text{g}}^{\tau-1}$ denotes the aggregated global model from the previous round $\tau - 1$.
Since the first round lacks a global model $\mathbf{w}_{\text{g}}^{\tau-1}$, we apply $m$-Krum~\cite{krum} on the local models, and select half of the local models to estimate a global model to prevent any negative impact from the malicious models. 
The algorithm then leverages the $\mathsf{L_2}$ scores to estimate a normal distribution and applies the $3\sigma$ rule to filter out potential malicious local models. 
Since models with lower \textit{evilness levels} are preferable, we apply a one-sided threshold of the $3\sigma$ rule: models with scores higher than $\mu + \lambda\sigma$ ($\lambda > 0$) are removed, while models with scores lower than $\mu - \lambda\sigma$ are retained.
The following theorem states that the likelihood of identifying a benign client as malicious decreases exponentially with $\lambda$.

\begin{algorithm}[ht]
    \caption{\name -Phase 2: Cross-Client Detection}
    \label{alg:cross_client}
    \textbf{Input:} $\tau$: training round ID ($\tau=0,1,\ldots$); $\mathcal{W}^\tau$: local models of round $\tau$; $m$: parameter of $m$-Krum; $\lambda$: parameter of $3\sigma$ Rule; $\mathbf{w}_g^{\mathit{ref}}$: global reference model from the previous round. 

    \begin{algorithmic}[1]
        \If{$\tau = 0$}~$m \gets |\mathcal{W}^\tau| / 2$, $f \gets |\mathcal{W}^\tau| / 2$, $\mathbf{w}_g^{\mathit{ref}} \gets \text{Krum\_and\_m\_Krum}(\mathcal{W}^\tau, m, f)$
        \EndIf

        \State $\mathcal{L} \leftarrow \text{compute}\_\text{L2}\_\text{scores}(\mathcal{W}^\tau, \mathbf{w}_g^{\mathit{ref}})$, $\mu \leftarrow \frac{\sum_{\ell\in \mathcal{L}} \ell}{|\mathcal{L}|}$, $\sigma \leftarrow \sqrt{\frac{\sum_{\ell\in \mathcal{L}} (\ell-\mu)^2}{|\mathcal{L}|-1}}$ \Comment{Estimate $\mathcal{N}$($\mu$, $\sigma$)}

        \ForAll{$\mathbf{w}_i \in \mathcal{W}^\tau$} 
            \If{$\mathcal{L}[i] > \mu + \lambda\sigma$}~Remove $\mathbf{w}_i$ from $\mathcal{W}^\tau$
            \EndIf
        \EndFor
        \State  Renew $\mathbf{w}_g^{\mathit{ref}}$ for the next round

        \State \Return $\mathcal{W}^\tau$ \Comment{Cache and return the filtered set}
    \end{algorithmic}
\end{algorithm}

 






\begin{theorem}\label{def: bound}
    Let $\mathcal{L}$ be the \emph{evilness level} scores for client models in the current FL round, where $\mathcal{L}$ follows normal distribution $\mathcal{N}$($\mu$, $\sigma$).
    The \emph{evilness level} for each client $i$ is computed as
    $\mathcal{L}[i]=||\mathbf{w}^\tau_i-\mathbf{w}^{\tau-1}_{\text{g}}||_2$. Under the Central Limit Theorem (CLT)~\cite{clt}, the probability that a benign client is erroneously flagged as malicious using the threshold $\mu + \lambda\sigma$ is upper bounded as $
        P(\mathcal{L}[i] > \mu + \lambda\sigma) \leq \frac{1}{\sqrt{2\pi}\lambda}e^{-\lambda^2/2}.$
\end{theorem}

\begin{proof}
    Let $\mathbf{w}_i \in \mathbb{R}^d$ be the local model parameters of client $i$, and $\mathbf{w}_{\text{g}} = \frac{1}{n}\sum_{j=1}^n \mathbf{w}_j$. Assume each parameter $w_{i,k}$ (for $k = 1, \ldots, d$) is a random variable with mean $\mu_k$ and variance $\sigma_k^2$. Due to CLT, for large $d$ (typical in ML models), the difference $\mathbf{w}_i - \mathbf{w}_{\text{g}}$ approximates a multivariate normal distribution $\mathcal{N}(0, \Sigma)$, where $\Sigma$ is the covariance matrix.
    The squared $\mathsf{L_2}$ norm $\|\mathbf{w}_i - \mathbf{w}_{\text{g}}\|_2^2$ follows a chi-squared distribution with $d$ degrees of freedom. For large $d$, this converges to $\mathcal{N}(d, 2d)$ by CLT. The square root ($\mathsf{L_2}$ norm) then approximates $\mathcal{N}(\sqrt{d - 1/2}, \sqrt{1/4})$ via the delta method.
    For $L[i] \sim \mathcal{N}(\mu, \sigma)$, the one-sided tail probability satisfies Mill's inequality $
    P(L[i] > \mu + \lambda\sigma) \leq \frac{1}{\sqrt{2\pi}\lambda} e^{-\lambda^2/2}$. Let $Z = \frac{L[i] - \mu}{\sigma} \sim \mathcal{N}(0, 1)$. Then $P(Z > \lambda) = \int_\lambda^\infty \frac{1}{\sqrt{2\pi}} e^{-z^2/2} dz \leq \frac{1}{\sqrt{2\pi}\lambda} e^{-\lambda^2/2}$,
    where the inequality follows from the bound $\int_\lambda^\infty e^{-z^2/2} dz \leq \frac{1}{\lambda} e^{-\lambda^2/2}$.
\end{proof}

\noindent\textbf{Effectiveness of the 3$\sigma$ Rule. } 
The effectiveness of the 3$\sigma$ rule in identifying malicious models is supported both theoretically and empirically for the following reasons:
\textit{i}) When client datasets are i.i.d., the parameters of local models are known to follow a normal distribution~\cite{baruch2019little, chen2017distributed, yin2018byzantine};
\textit{ii}) Even under non-i.i.d. settings, the Central Limit Theorem (CLT)~\cite{clt} ensures that local models tend to approximate a normal distribution, especially when the number of clients is at least 30~\cite{chang2006determination, CLT30};
\textit{iii}) Even when CLT does not hold strongly (\textit{e}.\textit{g}., the number of clients is lower than 30), prior work~\cite{karimireddy2020byzantine, rfa} shows that local models still exhibit certain statistical features, enabling the 3$\sigma$ rule to remain effective.
Furthermore, our empirical evaluations in \S\ref{sec:experiments} confirm the reliability of the 3$\sigma$ rule even with a small number of clients. This is because:
\textit{a}) SGD introduces noise during local training, which often causes model parameters to approximate normality in practice, even for a small number of clients; and
\textit{b}) The \textit{evilness level} of each local model aggregates high-dimensional model parameters, smoothing out individual irregularities and leading to a distribution that empirically resembles a Gaussian distribution.

\subsection{Extensions of \name}

\subsubsection{Optimization with importance layers} 

To perform the detection efficiently, in both the cross-round detection and the cross-client detection,
\name relies on \emph{importance layers}~\cite{foolsgold} of models, \textit{i}.\textit{e}., segmental representations of models rather than full model parameters, as references of full models in computation.
Specifically, it employs the second-to-last layer, as it retains substantial model information.  
An importance layer must satisfy:
\textit{i}) \textit{Representativeness}: capturing sufficient model information with minimal size (ideally a single layer of the original model), and
\textit{ii}) \textit{Generalizability}: applicability across diverse data distributions and model architectures.
We note that the importance layer is not required to contain the maximal information compared with other layers, but should be more \emph{informative} than the majority of the other layers. To improve the efficiency of computations, we select the second-to-last layer as the importance layer, as it retains substantial model information.
We experimentally validate the effectiveness of importance layer in \textbf{Exp 1} in \S\ref{sec:experiments}.

\subsubsection{Extensions against adaptive attacks}
To extend \name to be robust against adaptive attacks, cached global models from previous FL rounds cannot be used, as the global model is distributed to all clients and can be exploited by malicious participants. 
To address this issue, 
we adapt \name to operate without relying on cached models.
At the end of each round, the server estimates a global model using $m$-Krum~\cite{krum}, computed from the local models submitted in the current round, and employs it as a reference for both cross-round and cross-client detection. To further improve detection accuracy, we use the full model parameters instead of the importance layer. The algorithms are detailed as follows.

\paragraph{Cross-Round Detection}To identify suspicious activities of each FL training round, the cross-round detection estimates a global model with $m$-Krum (Algorithm~\ref{alg:krum}), where $m$ is set to half of the number of local models in the current round. 
The estimated global provides a reference for convergence without relying on global models of previous rounds; local models that deviate significantly from it are flagged as potentially malicious. Also, this estimated global model is used only for identifying potential attacks and does not impact the actual removal of models, thus is sufficient for providing a dependable reference.

\begin{algorithm}[ht]
    \caption{\name -Phase 1: Cross-Round Detection Against Adaptive Attacks}
    \label{algo:cross_round_adaptive}
    \textbf{Input:} $\tau$: training round ID ($\tau\geq 0)$; $\mathcal{W}^{\tau}$: client models of round $\tau$; $\gamma$: similarity threshold
    \begin{algorithmic}[1]
        \If{$\tau=0$}~\Return \textbf{True} \Comment{No previous models, activate cross-client detection by default}
        \EndIf
        
        \State 
        $\mathcal{W}^\tau \gets \text{clip}(\mathcal{W}^\tau)$ 
        \State $\mathbf{w}_g^{\text{ref}} \gets \text{Krum\_and\_m\_Krum}(\mathcal{W}^{\tau}, \frac{|\mathcal{W}^{\tau}|}{2}, \frac{|\mathcal{W}^{\tau}|}{2})$
        
        \ForAll{$\mathbf{w}_i^{\tau} \in \mathcal{W}^{\tau}$}
            \State 
            $\mathcal{S}_c(\mathbf{w}_g^{\text{ref}}, 
            \mathbf{w}_i^{\tau}) \gets \text{get\_cosine\_similarity}(\mathbf{w}_g^{\text{ref}}, \mathbf{w}_i^{\tau})$
            \If{$\mathcal{S}_c(\mathbf{w}_g^{\text{ref}}, \mathbf{w}_i^{\tau}) < \gamma$ 
            }
            \State \Return \textbf{True} \Comment{Potential attacks detected}
            \EndIf
        \EndFor
        \State \Return \textbf{False} \Comment{No attack detected}
    \end{algorithmic}
\end{algorithm}

\begin{algorithm}[ht]
    \caption{\name -Phase 2: Cross-Client Detection Against Adaptive Attacks}
    \label{alg:cross_client_adaptive}
    \textbf{Input:} $\tau$: training round ID ($\tau=0,1,\ldots$); $\mathcal{W}^\tau$: local models of round $\tau$; $m$: parameter of $m$-Krum ($m=|\mathcal{L}|/2$ by default); $\lambda$: parameter of the $3\sigma$ rule; $\mathbf{w}_g^{\mathit{ref}}$: global reference model from the last FL training round. 

    \begin{algorithmic}[1]
        \State $\mathbf{w}_g^{\mathit{ref}} \gets \text{get\_global\_model\_from\_cross\_round\_check}()$

        \State $\mathcal{L} \leftarrow \text{compute}\_\text{L2}\_\text{scores}(\mathcal{W}^\tau, \mathbf{w}_g^{\mathit{ref}})$
        \State $\mu \leftarrow \frac{\sum_{\ell\in \mathcal{L}} \ell}{|\mathcal{L}|}$, $\sigma \leftarrow \sqrt{\frac{\sum_{\ell\in \mathcal{L}} (\ell-\mu)^2}{|\mathcal{L}|-1}}$ \Comment{Estimate $\mathcal{N}$($\mu$, $\sigma$)}

        \ForAll{$\mathbf{w}_i \in \mathcal{W}^\tau$} \Comment{Operate on the original models $\mathcal{W}^\tau$, not the importance layers 
        }
            \If{$\mathcal{L}[i] > \mu + \lambda\sigma$}
            \State Remove $\mathbf{w}_i$ from $\mathcal{W}^\tau$
            \EndIf
        \EndFor
        \State \Return $\mathcal{W}^\tau$ \Comment{Cache and return the filtered set}
    \end{algorithmic}
\end{algorithm}

The algorithm is summarized in Algorithm~\ref{algo:cross_round_adaptive}. 
Upon the server receiving local models from clients, it clips the models based on a randomly selected norm, and applies $m$-Krum~\cite{krum} to computes an estimated global model $\mathbf{w}_g^\mathit{ref}$. 
Then, it computes cosine similarities between each local model and $\mathbf{w}_g^\mathit{ref}$. 
Local models exhibiting higher similarities to these reference models are deemed more likely to be benign, while those with lower similarities are considered suspicious, activating a  rigorous cross-client detection in the next phase.
We note that \name just flags suspicious models without removing them, thus, \name does not rely heavily on cosine similarities.

\paragraph{Cross-Client Detection}
To extend the cross-client detection against adaptive attacks, we modify it to use the estimated global model $\mathbf{w}_g^\mathit{ref}$ from the cross-round detection as a reference instead of the cached global model from the last FL round. The modified algorithm is summarized in Algorithm~\ref{alg:cross_client_adaptive}.

\subsubsection{Extensions to client sampling}
For ease of explanation, we assume that all clients participate in aggregation in every FL round. With some engineering efforts, our method can be extended with client selection. We can cache historical client models for the same clients across rounds, such that the server can perform cross-round detection even when clients do not participate in every round. If the cached model for a client is too old, we can use the global model from the last round as the reference model. A scenario with adversary clients that participate only once (\textit{i}.\textit{e}., single-shot attacks) constitutes a specific case of the client selection challenge described above. In such cases, we can use the global model from the last round as the reference model for cross-round detection.

\section{Verifiable Anomaly Detection}\label{sec:zkp}

\begin{figure*}
    \centering
    \includegraphics[width=0.98\textwidth]{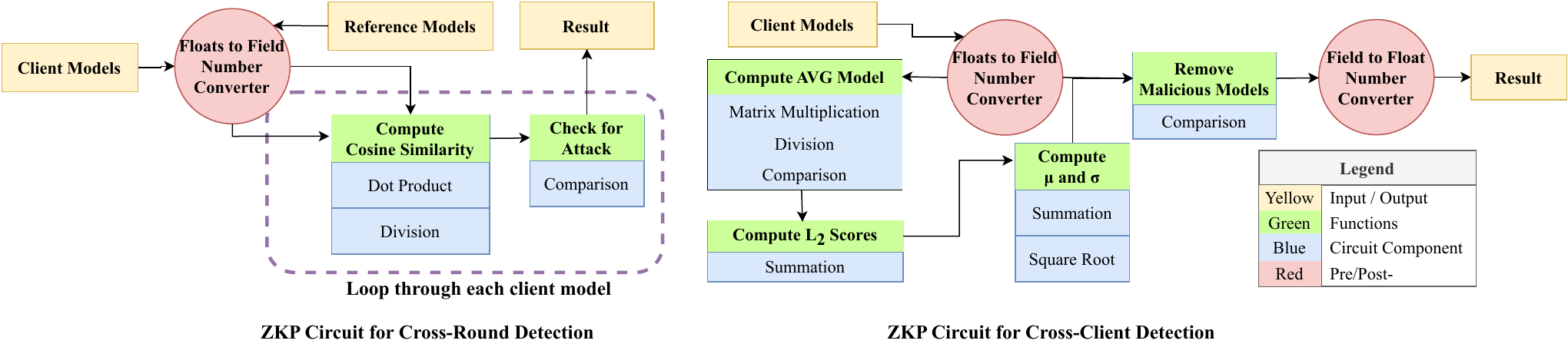}
    \caption{ZKP circuits designed for \name.}
    \label{fig:ZKP_circuit}
\end{figure*}

In FL systems with a defense mechanism, a critical trust gap arises as clients cannot  verify the server’s honest execution of the defense mechanism, forcing them to rely on the server’s integrity.
To address this issue, we integrate ZKPs that enable a prover (\textit{i}.\textit{e}., the FL server) to demonstrate computational correctness to the verifiers (\textit{i}.\textit{e}., clients) without revealing sensitive inputs, \textit{e}.\textit{g}., client models or detection thresholds. 

ZKPs bridge the trust gap in the FL systems with two critical properties:
\textit{i}) \textit{Client-Side Verification:} clients can independently verify that the mechanism has been executed faithfully, without trusting the server blindly; and \textit{ii}) \textit{Privacy Preservation:} verification does not require exposing private data, such as other clients’ local models or internal server parameters, ensuring confidentiality while maintaining system integrity. 

\subsection{ZKP Circuit Design}

We design ZKP circuits as in~\cref{fig:ZKP_circuit}. 
Computations in the two detection stages are linear and can be compiled into an arithmetic circuit easily, \textit{e}.\textit{g}., computing cosine similarity between two matrices of size $n\times n$ requires a circuit with $O(n^2)$ multiplication gates and one division. While computing division on a circuit directly is difficult, we can verify with the prover easily, providing the pre-computed quotient and remainder beforehand. 
Our key optimizations are as follows:

\textbf{\textit{i}) Freivalds’ algorithm~\cite{freivalds1977probabilistic}:} We leverage Freivalds' algorithm~\cite{freivalds1977probabilistic,weng2021mystique} to verify matrix multiplications. 
In general, the matrix multiplication constitutes the basis of the verification schemes in \name. Naively verifying a matrix multiplication $AB = C$ where $A,B,C$ are of size $n\times n$ requires proving the computation step by step, which requires $O(n^3)$ multiplication gates. With Freivalds' algorithm, the prover first computes the result off-circuit and commits to it. Then, the verifier generates a random vector $v$ of length $n$, and checks $A(Bv)\stackrel{?}{=}Cv$. This approach reduces the circuit size to $O(n^2)$.

\textbf{\textit{ii}) Approximate square roots:} To verify that $x = \sqrt{y}$ is computed correctly, we ask the prover to provide the answer $x$ as witness and then we check in the ZKP that $x$ is indeed the square root of $y$. Note that we cannot check $x^2$ is equal to $y$ because the zkSNARK works over a prime field and the square root of an input number might not exist. So, we check if $x^2$ is close to $y$ by checking that $x^2 \le y$ and $(x+1)^2 \ge y$. This approach reduces the computation of square root to 2 multiplications and 2 comparisons.

{The zero-knowledge property of ZKPs allows public verification of prover’s (\textit{i}.\textit{e}., the FL server) integrity in case of the server being untrusted. 
By incorporating ZKPs, we provide a public verifiable approach for each client to ensure the FL server's integrity, which is crucial for establishing and maintaining trust in FL systems. This allows the FL clients to verify the correctness of the defense without relying solely on the server's goodwill. 
Moreover, the approach remains secure in the presence of adversarial clients, as ZKP reveals no information about the prover’s witness—\textit{i}.\textit{e}., the server’s private data, models, or thresholds used in the method.
}

\subsection{ZKP-compatible language.}
The first challenge of applying ZKP protocols is to convert the computations into a ZKP-compatible language. ZKP protocols model computations as arithmetic circuits with addition and multiplication gates over a prime field. However, our computations for our approach are over real numbers. The second challenge is that some computations such as square root are nonlinear, making it difficult to wire them as a circuit. To address these issues, we implement a class of operations that map real numbers to fixed-point numbers. To build our ZKP scheme, we use Circom library \cite{circom}, which compiles the description of an arithmetic circuit in a front-end language similar to C++ to the back-end ZKP protocol.

\subsection{ZKP implementation}

To implement ZKP in FL systems, we employ zkSNARKs~\cite{zkp}, a ZKP variant with constant proof size and verification time, regardless of the size of computation. It is essential for real deployments, where clients often run under resource constraints. Our design (shown  in~\cref{fig:ZKP_circuit}) ensures that verification overhead remains minimal, even for large-scale computations. 
We note that the computations in Algorithms~\ref{algo:cross_round} and~\ref{alg:cross_client} rely heavily on linear operations, which we translate into arithmetic circuits for ZKP compatibility. For instance, computing cosine similarity between two $n\times n$
matrices requires 
$\mathcal{O}(n^2)$ multiplication gates and a division operation. While division is non-trivial, we circumvent this by having the prover precompute quotients and remainders, which the circuit verifies via modular arithmetic.

In our implementation, we use the Groth16~\cite{groth2016size} zkSNARK scheme implemented in the Circom library~\cite{circom} for all the computations described earlier. We choose this ZKP scheme because its construction ensures constant proof size (128 bytes) and constant verification time. Because of this, Groth16 is popular for blockchain applications as it necessitates little on-chain computation. There are other ZKP schemes based on different constructions that can achieve faster prover time~\cite{liu2021zkcnn}, but their proof size is bigger and verification time is not constant, which is a problem if the verifier lacks computational power, as in our case since the verifiers are the FL clients in our setting. The construction of a ZKP scheme that is efficient for both the prover and verifier is still an open research direction.

\textit{Interactivity of zkSNARKs. }
In the Freivalds’ algorithm~\cite{freivalds1977probabilistic}, the prover first computes the matrix multiplication and commits to its result. Then the verifier generates and sends the random vector. This step is interactive in nature, but we can make this non-interactive using the Fiat-Shamir heuristic \cite{fiat1986prove} as it is public-coin, meaning the vector is randomly selected by the verifier and made public to everyone. Therefore, the prover can instead generate this vector by setting it to the hash of matrices $A$, $B$ and $C$. With this, our entire ZKP pipeline, including the Freivalds’ step can become truly non-interactive.

\begin{figure*}[htbp]
    \centering
    \begin{subfigure}[t]{0.248\textwidth}
        \centering
        \includegraphics[width=\linewidth]{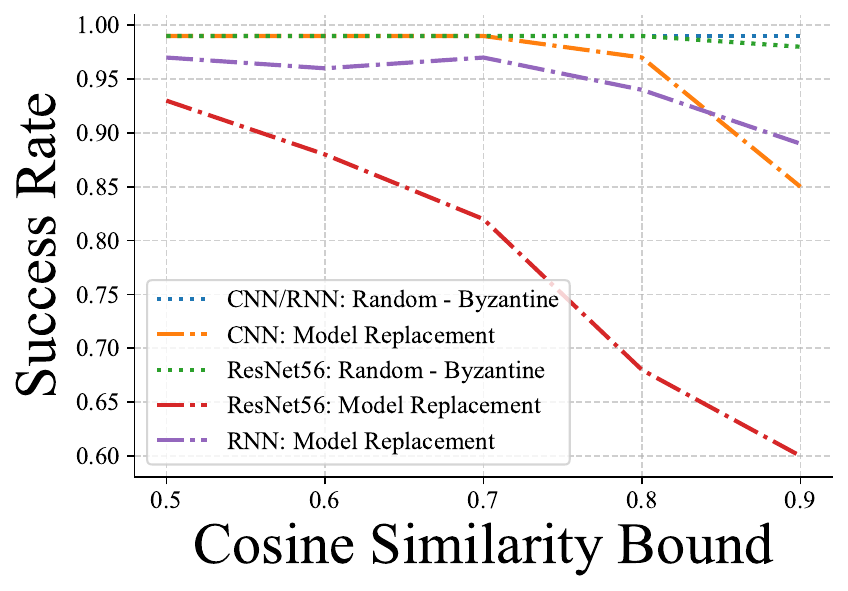}
        \caption{Varying $\gamma$.}
        \label{fig:phase1_similarity_bound}
    \end{subfigure}\hfill
    \begin{subfigure}[t]{0.248\textwidth}
        \centering
        \includegraphics[width=\linewidth]{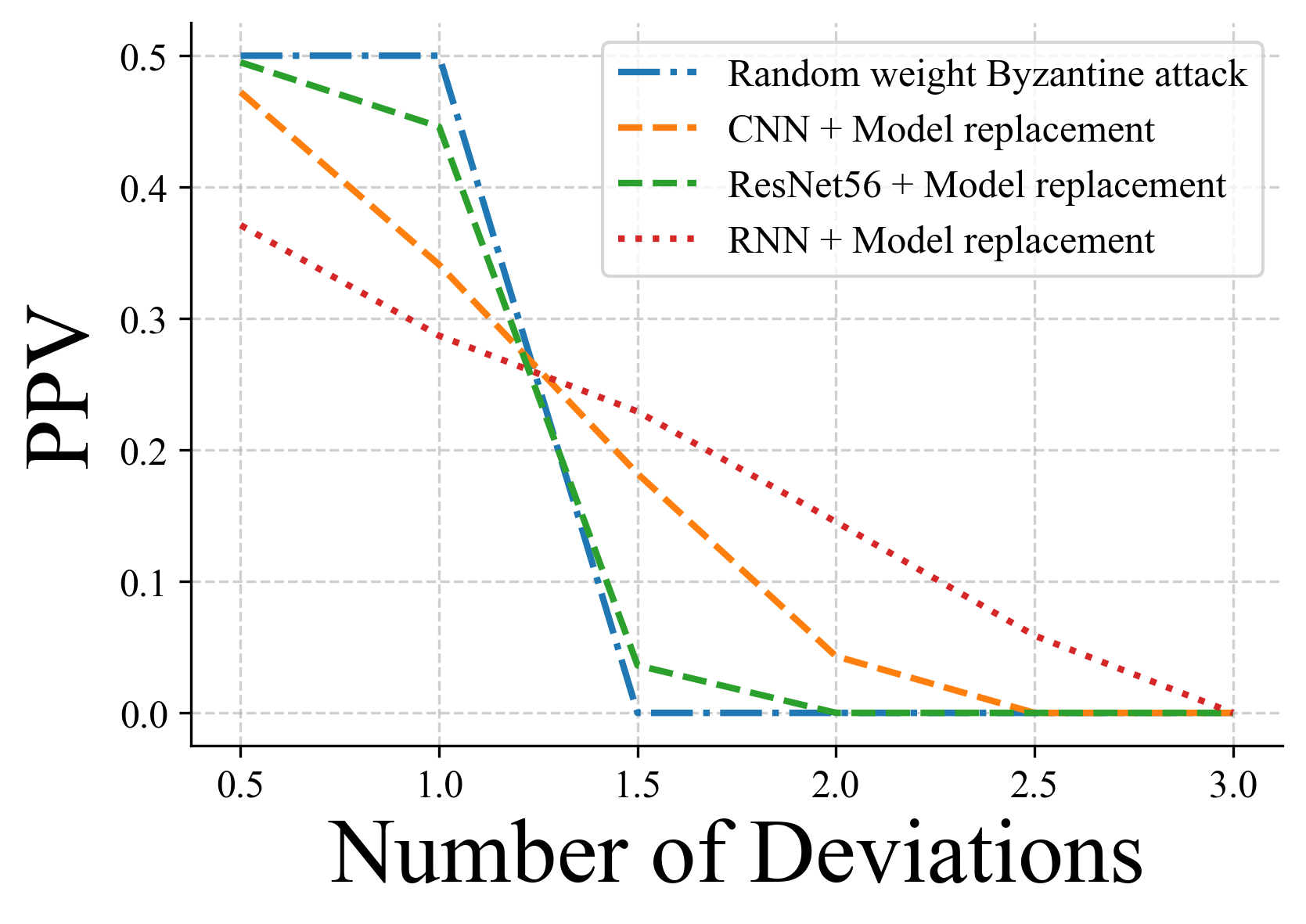}
        \caption{Varying \# deviations.}
        \label{fig:varying_dev_num}
    \end{subfigure}\hfill
    \begin{subfigure}[t]{0.248\textwidth}
        \centering
        \includegraphics[width=\linewidth]{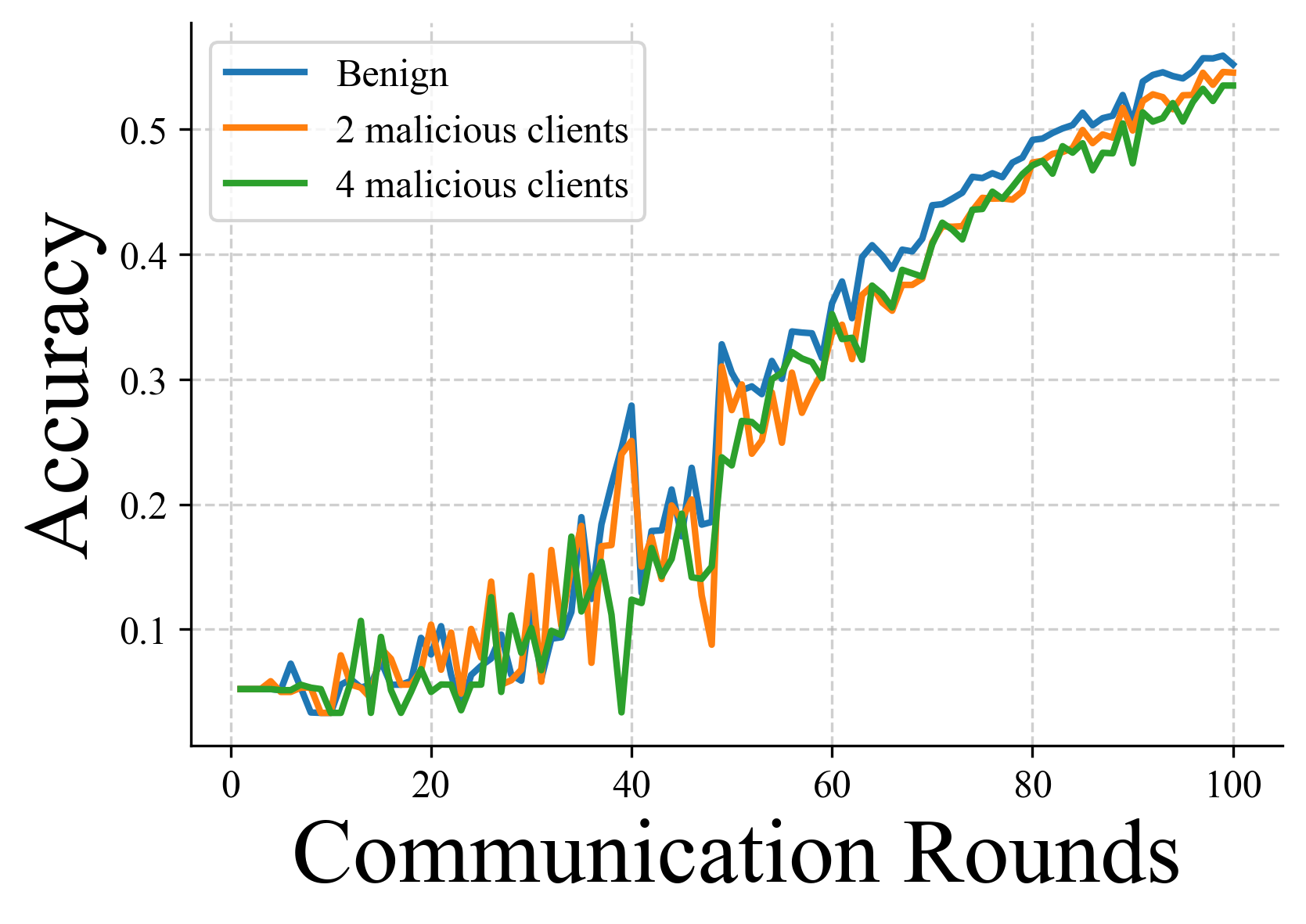}
        \caption{\# malicious clients.}
        \label{fig:malicious_num_exp}
    \end{subfigure}\hfill
    \begin{subfigure}[t]{0.248\textwidth}
        \centering
        \includegraphics[width=\linewidth]{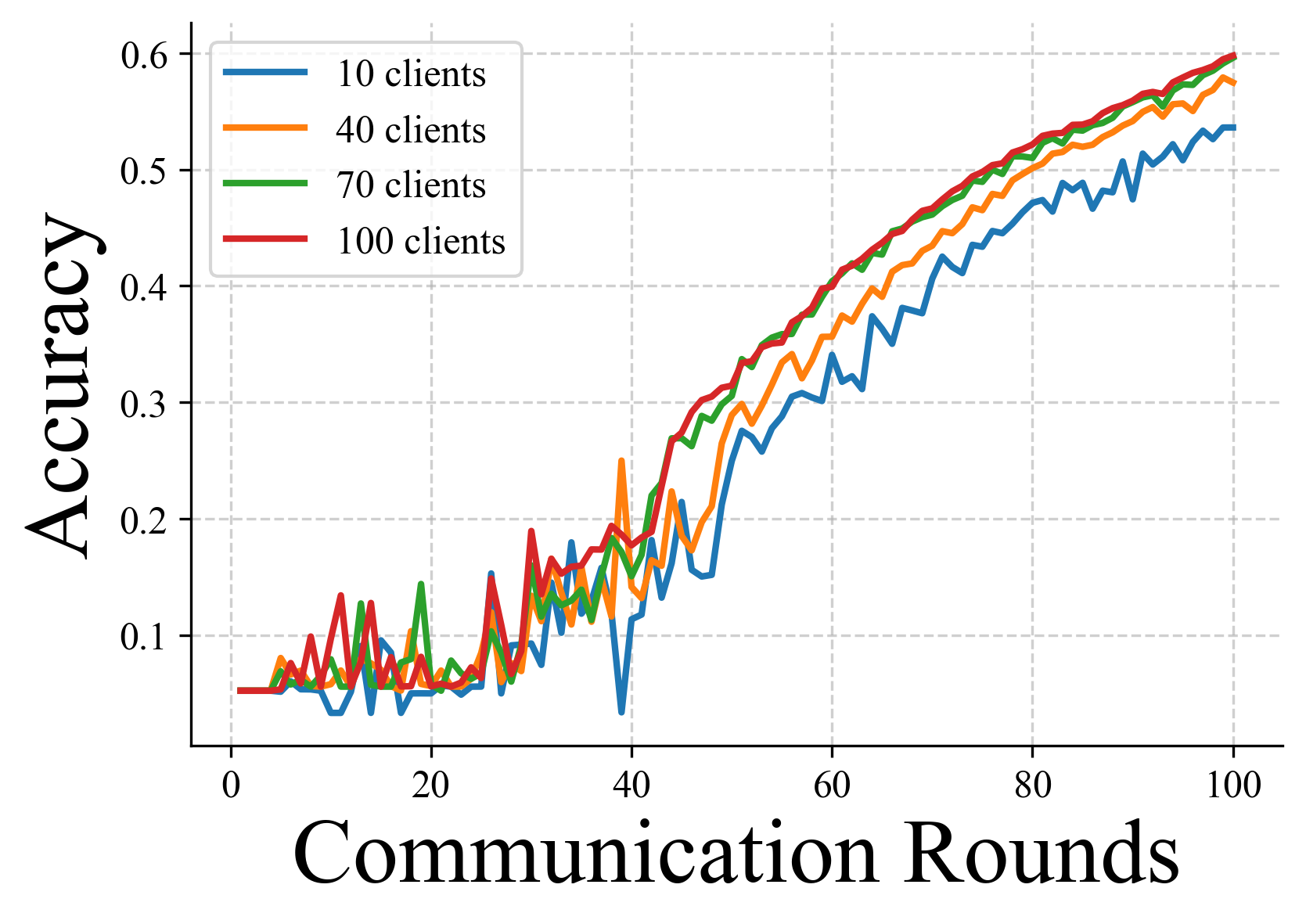}
        \caption{Varying \# clients.}
        \label{fig:vary_client_num}
    \end{subfigure}
    \caption{Impacts of different parameters.}
    \label{fig:parameter_impact}
\end{figure*}

\begin{figure*}[htbp]
    \centering
    \begin{minipage}[t]{0.496\textwidth} 
        \centering
        \begin{subfigure}[t]{0.496\textwidth}
            \centering
            \includegraphics[width=\linewidth]{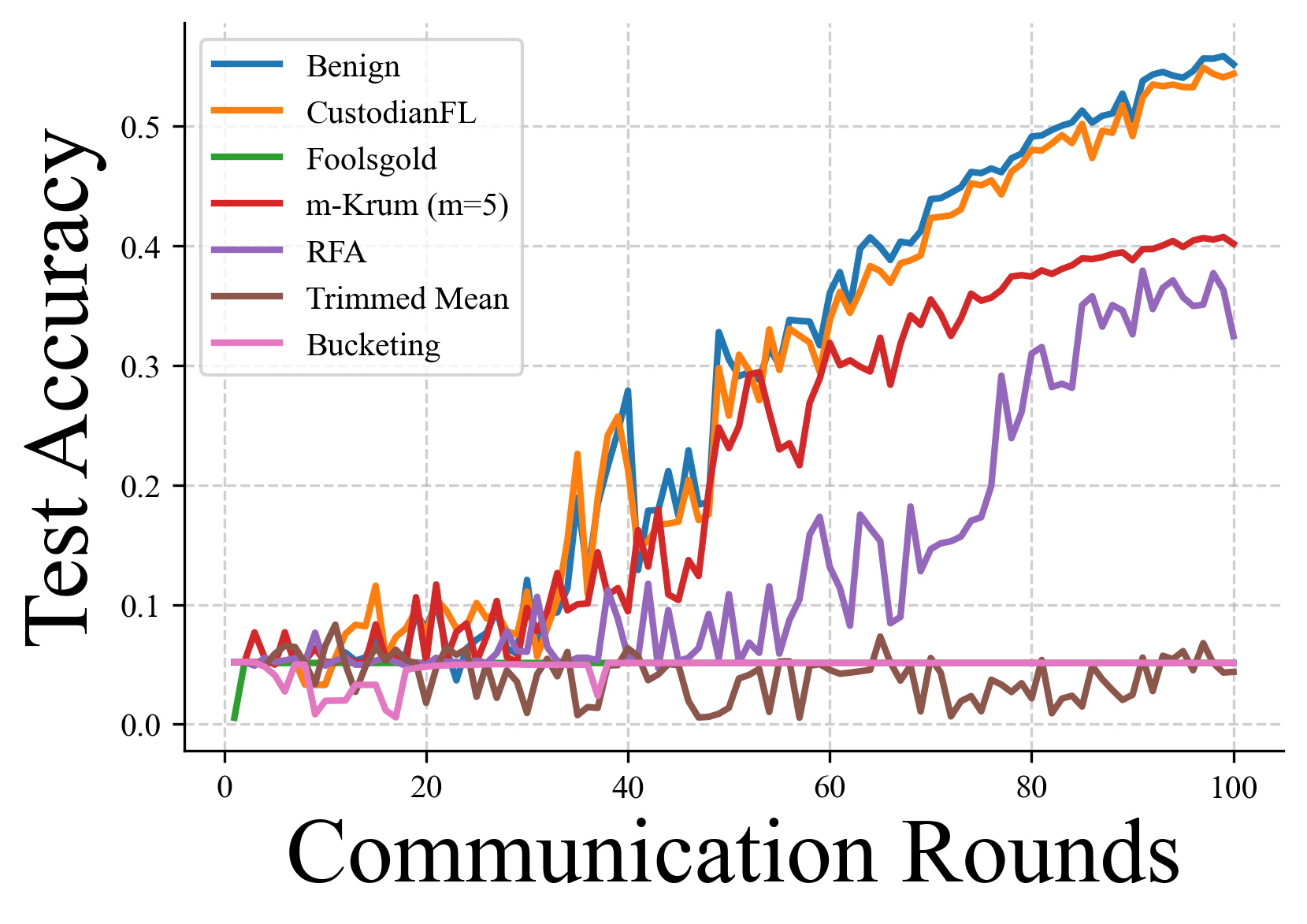}
            \caption{Random weights.}
            \label{fig:compare_defenses_random_byzantine}
        \end{subfigure}\hfill
        \begin{subfigure}[t]{0.496\textwidth}
            \centering
            \includegraphics[width=\linewidth]{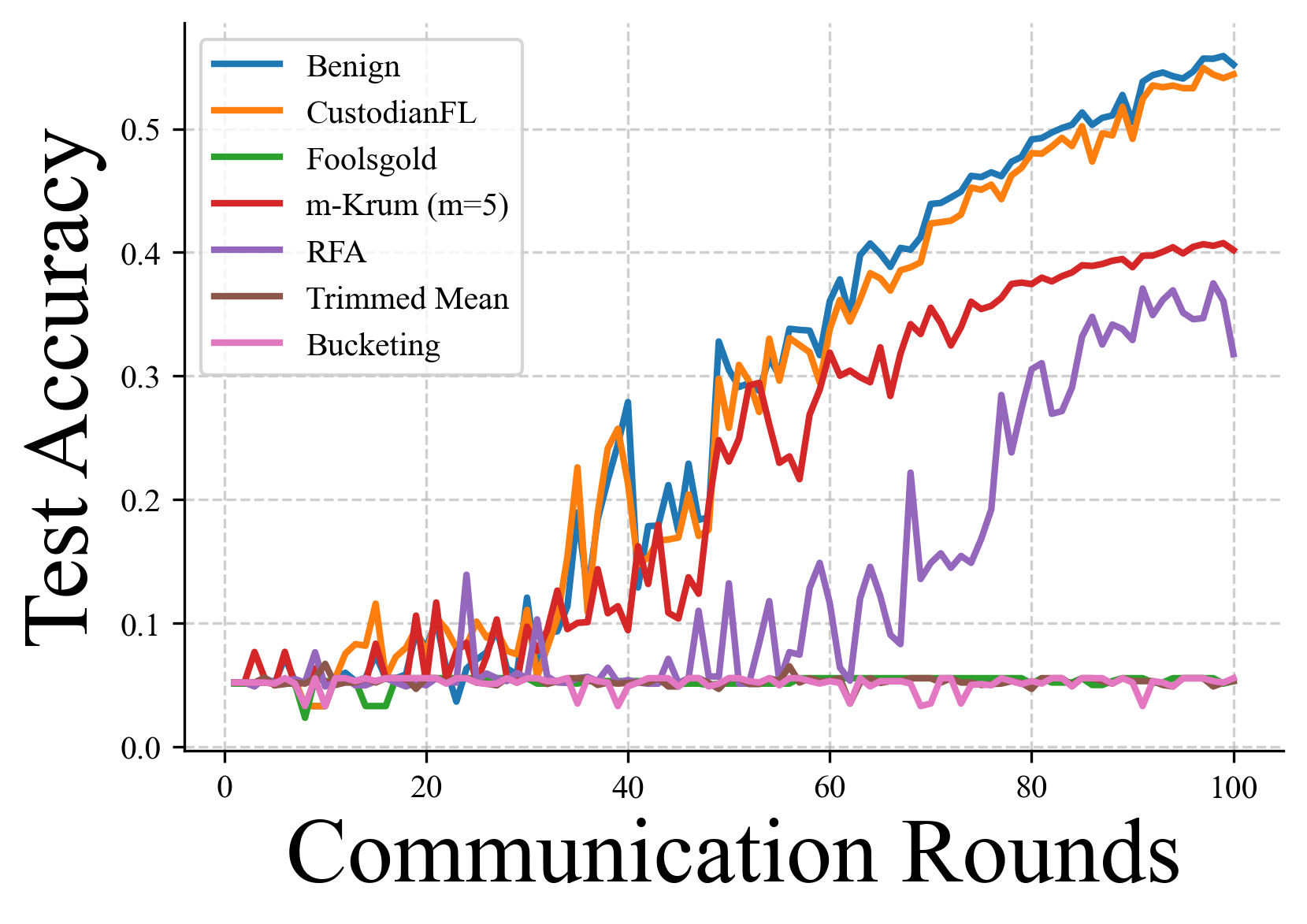}
            \caption{Zero weights.}
            \label{fig:byzantine_zero}
        \end{subfigure}
        \caption{Byzantine attacks.}
    \end{minipage}
    \hfill
    \begin{minipage}[t]{0.496\textwidth} 
        \centering
        \begin{subfigure}[t]{0.496\textwidth}
            \centering
            \includegraphics[width=\linewidth]{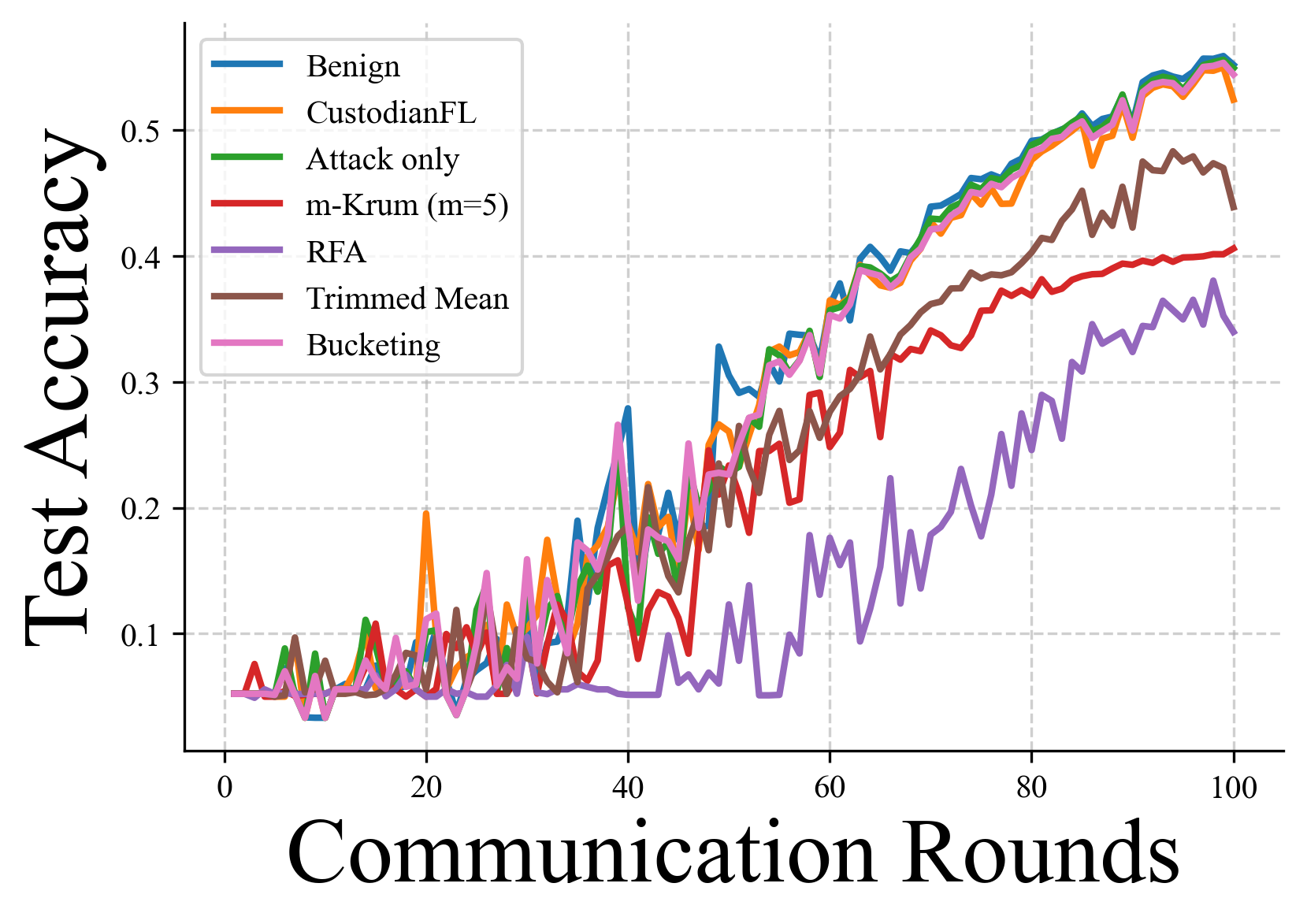}
            \caption{Label flipping.}
            \label{fig:label_flipping_exp}
        \end{subfigure}\hfill
        \begin{subfigure}[t]{0.496\textwidth}
            \centering
            \includegraphics[width=\linewidth]{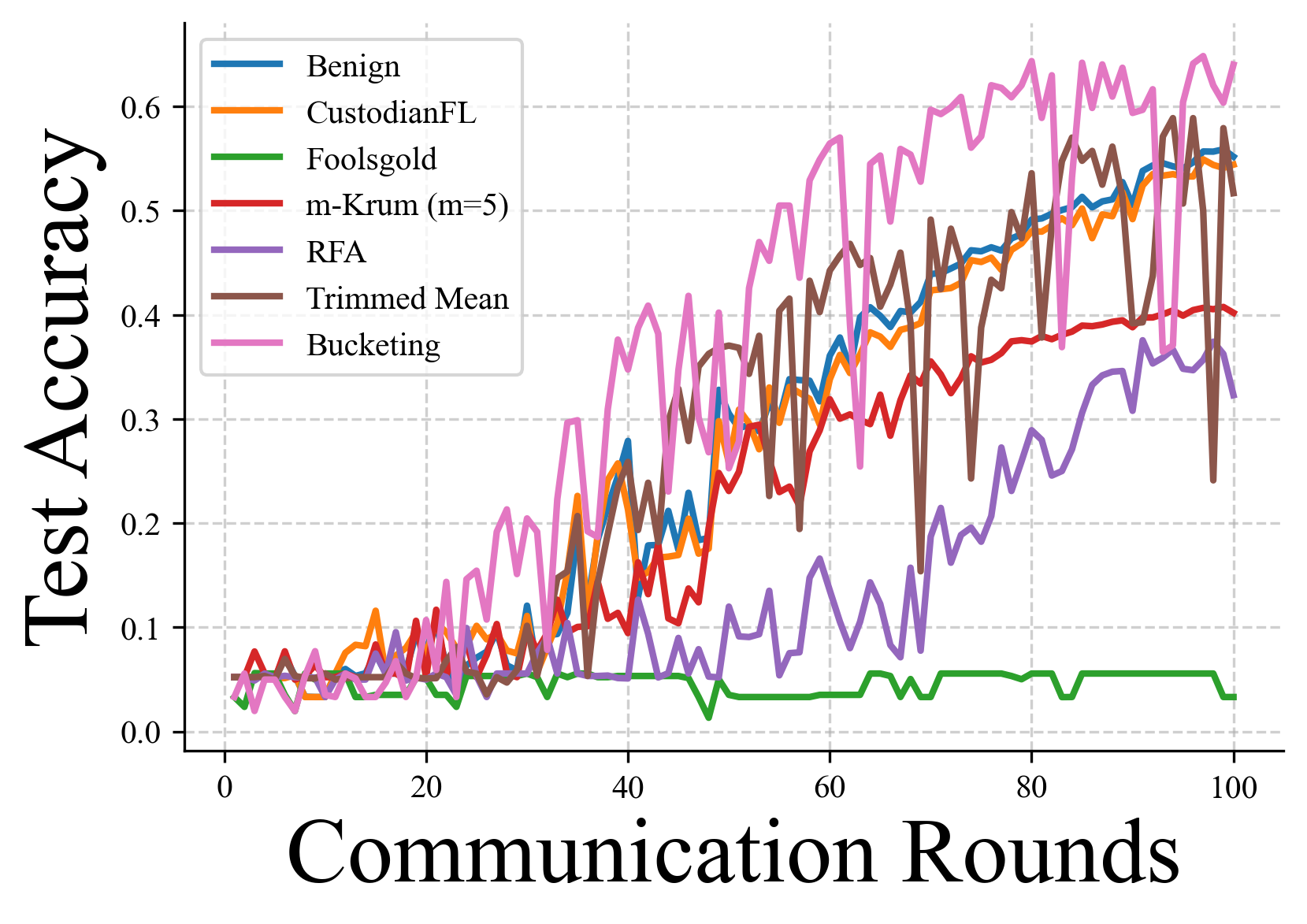}
            \caption{Model replacement.}
            \label{fig:model_replacement}
        \end{subfigure}
        \caption{Backdoor attacks.}
    \end{minipage}
\end{figure*}

\section{Evaluations}\label{sec:experiments}

\noindent\textbf{Models and Datasets. }We evaluate \name across a diverse set of model–dataset pairs that are widely used in FL research. For CV tasks, we adopt CNN~\cite{mcmahan2017communication} on the FEMNIST dataset~\cite{caldas2018leaf}, ResNet-20~\cite{he2016deep} on CIFAR-10~\cite{cifar}, and ResNet-56~\cite{he2016deep} on CIFAR-100~\cite{cifar}. For NLP tasks, we employ RNN~\cite{mcmahan2017communication} on the Shakespeare dataset~\cite{mcmahan2017communication}. 

\noindent\textbf{Setting.} By default, we employ CNN and the non-i.i.d. FEMNIST dataset ($\alpha = 0.5$), as the non-i.i.d. setting closely captures practical scenarios. training and the ``labelled'' subset for testing.
We utilize FedAVG in our experiments. We vary the number of clients from 10 to 100 in \textbf{Exp 5},
and by default, we use 10 clients for FL training, corresponding to practical FL applications where the number of clients is typically less than 10, especially in ToB scenarios. 
For evaluations on adaptive attacks, we leverage ResNet50 and CIFAR100, and set the proportion of malicious clients to 40\% by default. 
We implement the ZKP system in Circom~\cite{circom}.
We conduct our evaluations on a server with 8 NVIDIA A100-SXM4-80GB GPUs, and validate the correct execution with ZKP on Amazon AWS with an m5a.4xlarge instance with 16 CPU cores and 32 GB memory.

\noindent\textbf{Selection of attacks and defenses. }
We employ two Byzantine attacks and two backdoor attacks that are widely considered in the literature, including \textit{i}) a random weight Byzantine attack that randomly modifies the local submissions~\cite{chen2017distributed,fang2020local}, \textit{ii}) a zero weight Byzantine attack that sets all model weights to zero~\cite{chen2017distributed,fang2020local}, \textit{iii}) a label flipping backdoor attack that flips labels in the local data~\cite{tolpegin2020data}, and \textit{iv}) a model replacement backdoor attack~\cite{how_to_backdoor} that intends to use a poisoned local model to replace the global model. 
We utilize 5 baseline defense mechanisms that can be effective in real systems, including $m$-Krum~\cite{krum}, Foolsgold~\cite{foolsgold}, RFA~\cite{rfa}, Bucketing~\cite{karimireddy2020byzantine},  and Trimmed Mean~\cite{yin2018byzantine}. For $m$-Krum, by default, we set $m$ to 5, which means 5 out of 10 submitted local models participate in aggregation in each FL training round.
We test our method from the earliest stages of training (\textit{i}.\textit{e}., training from scratch), instead of after model convergence, to reflect practical FL scenarios where adversaries may attack at any point, including during initial model convergence. We do so because early-stage attacks are more challenging: benign local models can exhibit significant variability due to non-i.i.d. data distributions and random initialization. Such variability makes it inherently harder to distinguish malicious models from benign ones, creating a more rigorous testbed for defenses.

\noindent\textbf{Exp 1: Selection of importance layer.} {We utilize {the $\mathsf{L_2}$ norm of the local models }to evaluate the ``sensitivity'' of each layer.} A layer with a norm higher than most of the other layers indicates higher sensitivity compared to others, thus can be utilized to represent the whole model. The results for RNN, CNN, and ResNet-56 are deferred to~\cref{fig:rnn_importance_feature},~\cref{fig:cnn_importance_feature}, and~\cref{fig:resnet56_importance_feature} in Appendix, respectively. The results show the sensitivity of the second-to-the-last layer is higher than most of the other layers. Thus, this layer includes adequate information of the whole model and can be selected as the importance layer.

\noindent\textbf{Exp 2: Impact of the similarity threshold.} We evaluate the impact of the similarity threshold $\gamma$ in the cross-round check with 10 clients in each FL round, where 4 of them are malicious. Ideally, the cross-round check should confirm the absence or presence of an attack accurately. 
We evaluate the impact of the cosine similarity threshold $\gamma$ in the cross-round check by setting $\gamma$ to 0.5, 0.6, 0.7, 0.8, and 0.9. According to Algorithm~\ref{algo:cross_round}, a cosine score lower than $\gamma$ shows lower similarities between the local models and their reference models, indicating the potential occurrence of an attack. 
As described in~\cref{fig:phase1_similarity_bound}, the cross-round detection success rate is close to 100\% in the case of Byzantine attacks. We observe that, when the cosine similarity threshold $\gamma$ is set to 0.5, the performance is satisfactory in all cases, with at least 93\% cross-round detection success rate.

\noindent\textbf{Exp 3: Selection of the number of deviations ($\lambda$).} We vary $ \lambda$ to 0.5, 1, 1.5, 2, 2.5, and 3, and utilize $\textsf{PPV}$ to evaluate the impact of the number of deviations, \textit{i}.\textit{e}., the parameter $\lambda$ in the anomaly bound $\mu+\lambda\sigma$. To evaluate a challenging case where a large portion of the clients are malicious, we set 40\% clients as malicious in each FL round. Given that the number of FL rounds is 100, the total number of malicious submissions is 400. We 
evaluate our approach on three tasks, as follows: \textit{i}) CNN+FEMNIST, \textit{ii}) ResNet-56+Cifar100, and \textit{iii}) RNN + Shakespeare. We observe in~\cref{fig:varying_dev_num}, 
that when $\lambda$ is 0.5, the results are the best. 
Especially for the random weight Byzantine attack, we see that the $\textsf{PPV}$ is exactly 0.5, indicating that all malicious local models are detected. In subsequent experiments, unless specified otherwise, we set $\lambda$ to 0.5.

\begin{figure*}[htbp]
    \centering
    \begin{minipage}[t]{0.496\textwidth}
        \centering
        \begin{subfigure}[t]{0.496\textwidth}
            \centering
            \includegraphics[width=\linewidth]{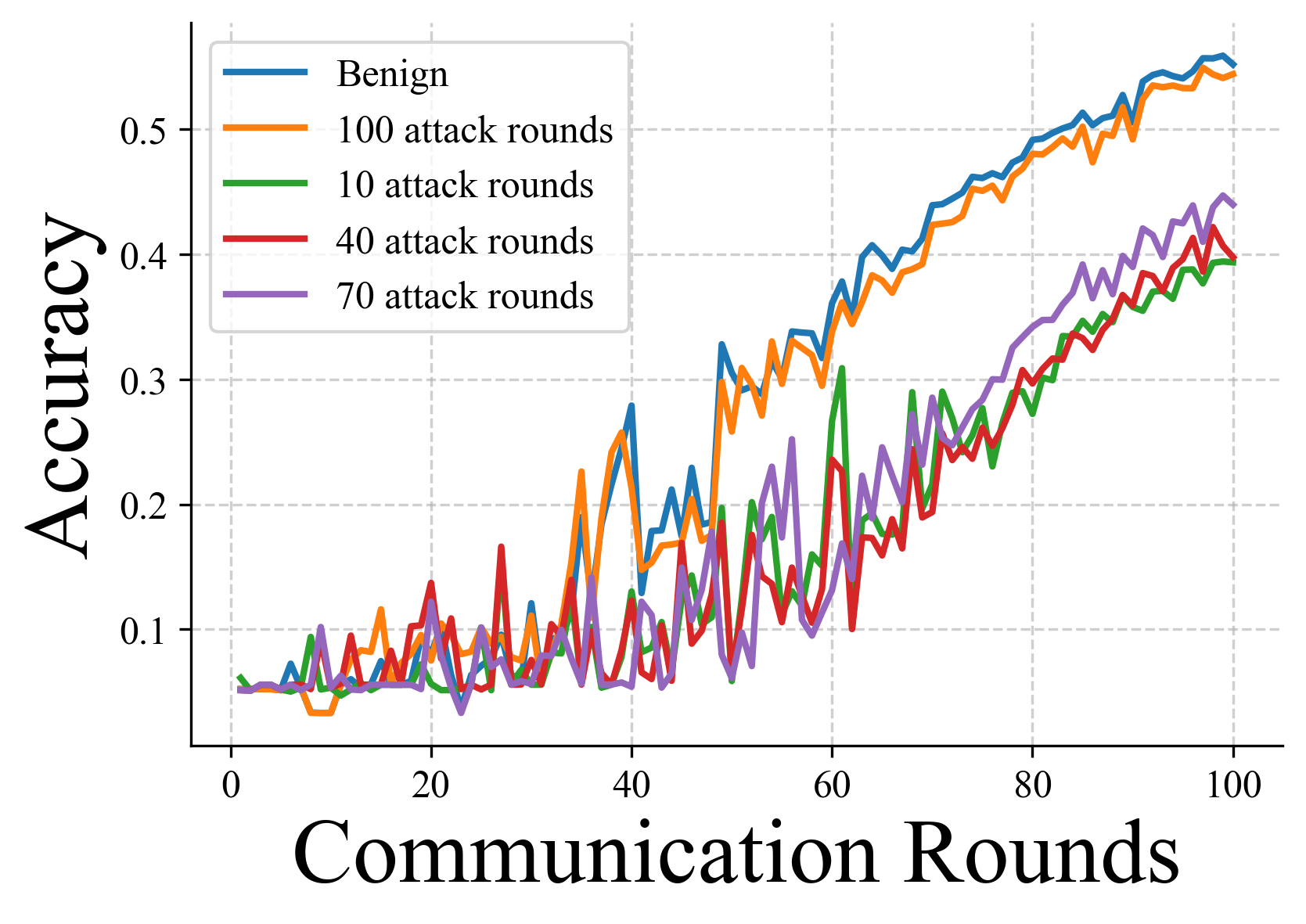}
            \caption{Varying \# attack rounds.}
            \label{fig:selected_attack}
        \end{subfigure}\hfill
        \begin{subfigure}[t]{0.496\textwidth}
            \centering
            \includegraphics[width=\linewidth]{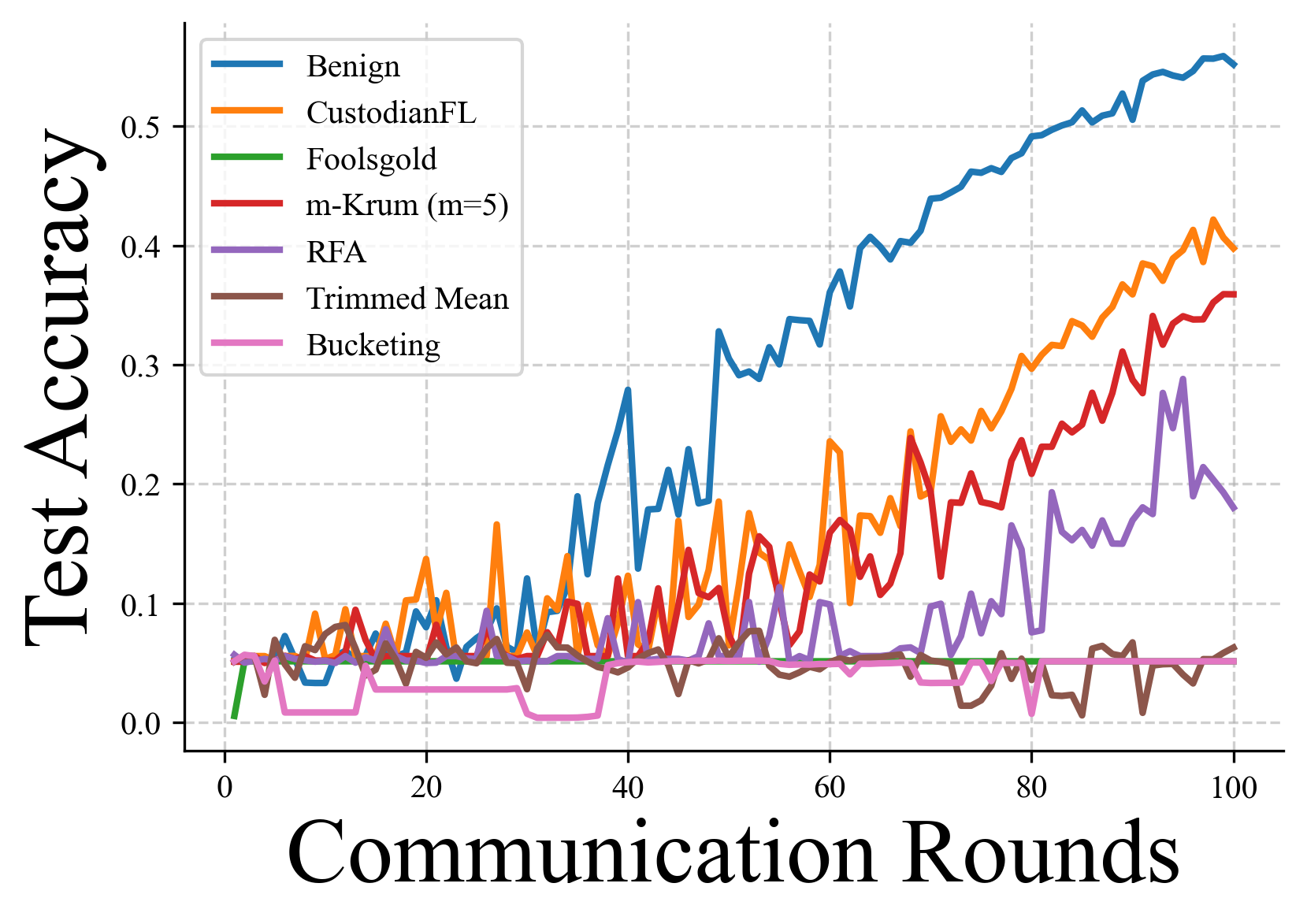}
            \caption{40 attack rounds.}
            \label{fig:40attack_rounds}
        \end{subfigure}
        \caption{Evaluations on selected attacks.}
        \label{fig:selected_attacks}
    \end{minipage}\hfill
    \begin{minipage}[t]{0.496\textwidth}
        \centering
        \begin{subfigure}[t]{0.496\textwidth}
            \centering
            \includegraphics[width=\linewidth]{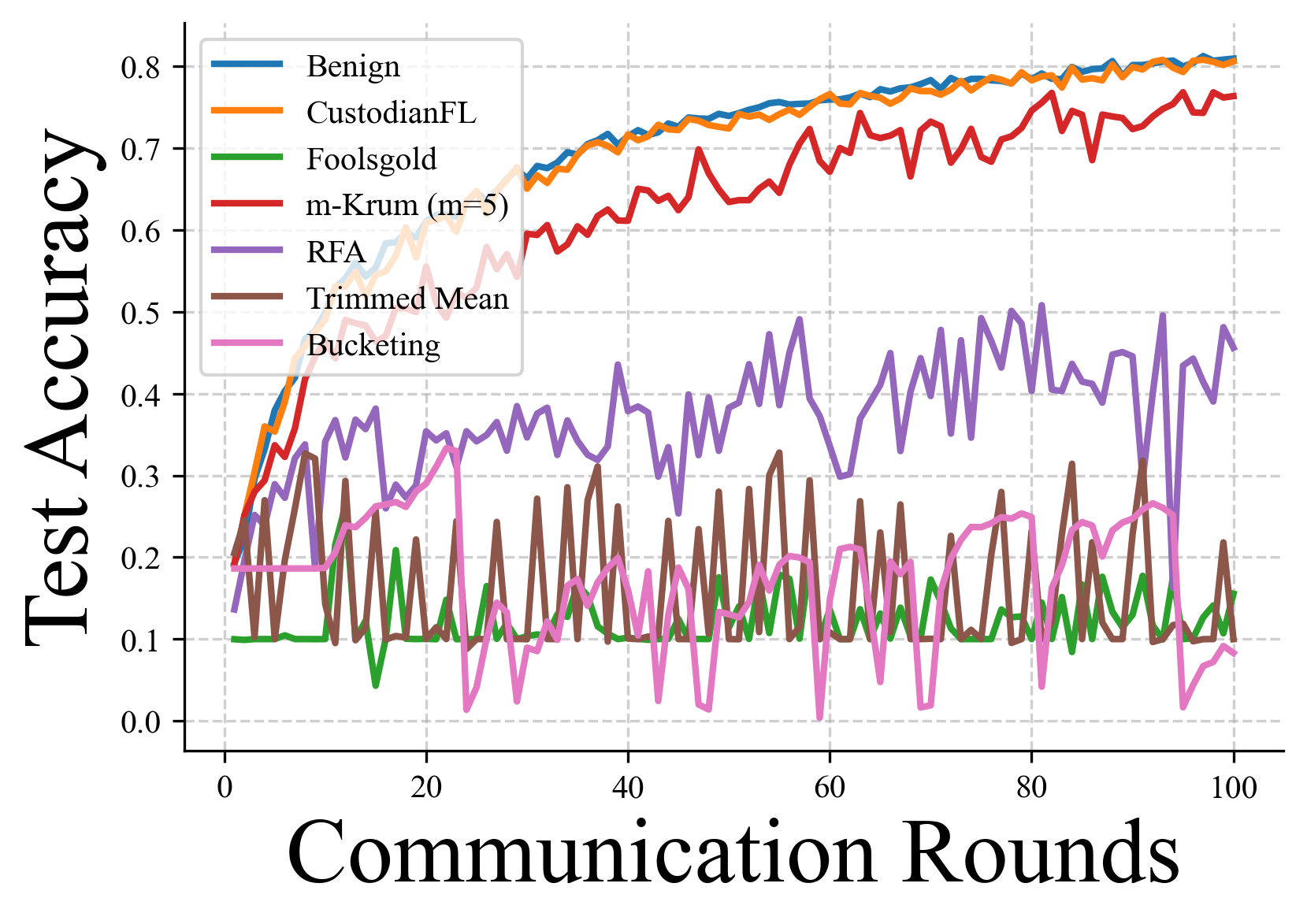}
            \caption{ResNet-20, CIFAR-10.}
            \label{fig:resnet20_exp}
        \end{subfigure}\hfill
        \begin{subfigure}[t]{0.496\textwidth} 
            \centering
            \includegraphics[width=\linewidth]{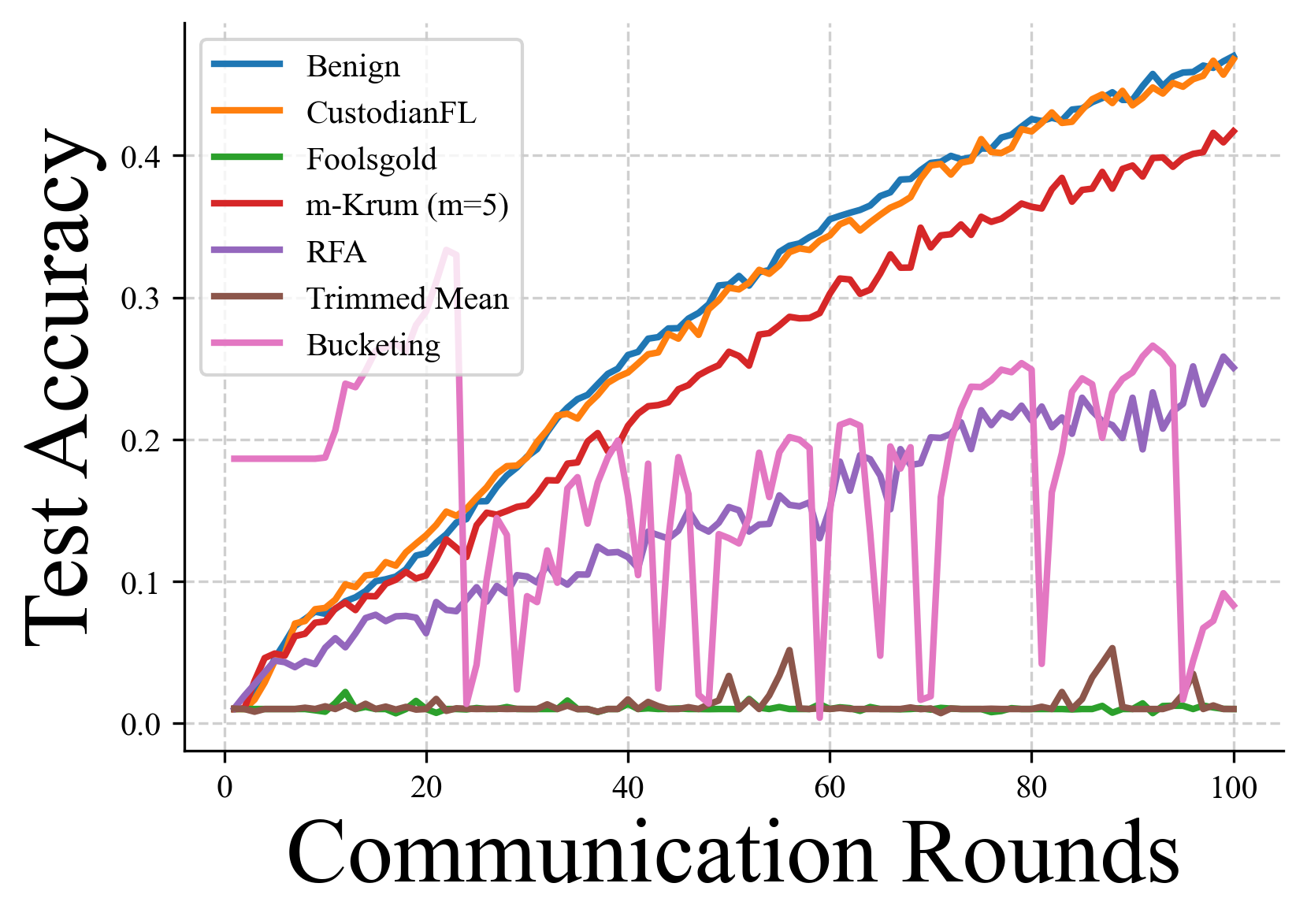}
            \caption{ResNet-56, CIFAR-100.}
            \label{fig:resnet56_exp}
        \end{subfigure}
        \caption{Evaluations on CV tasks.}
        \label{fig:cv_tasks}
    \end{minipage}
\end{figure*}

\begin{figure*}[htbp]
    \centering
    \begin{minipage}[t]{0.496\textwidth}
        \centering
        \begin{subfigure}[t]{0.496\textwidth}
            \centering
            \includegraphics[width=\linewidth]{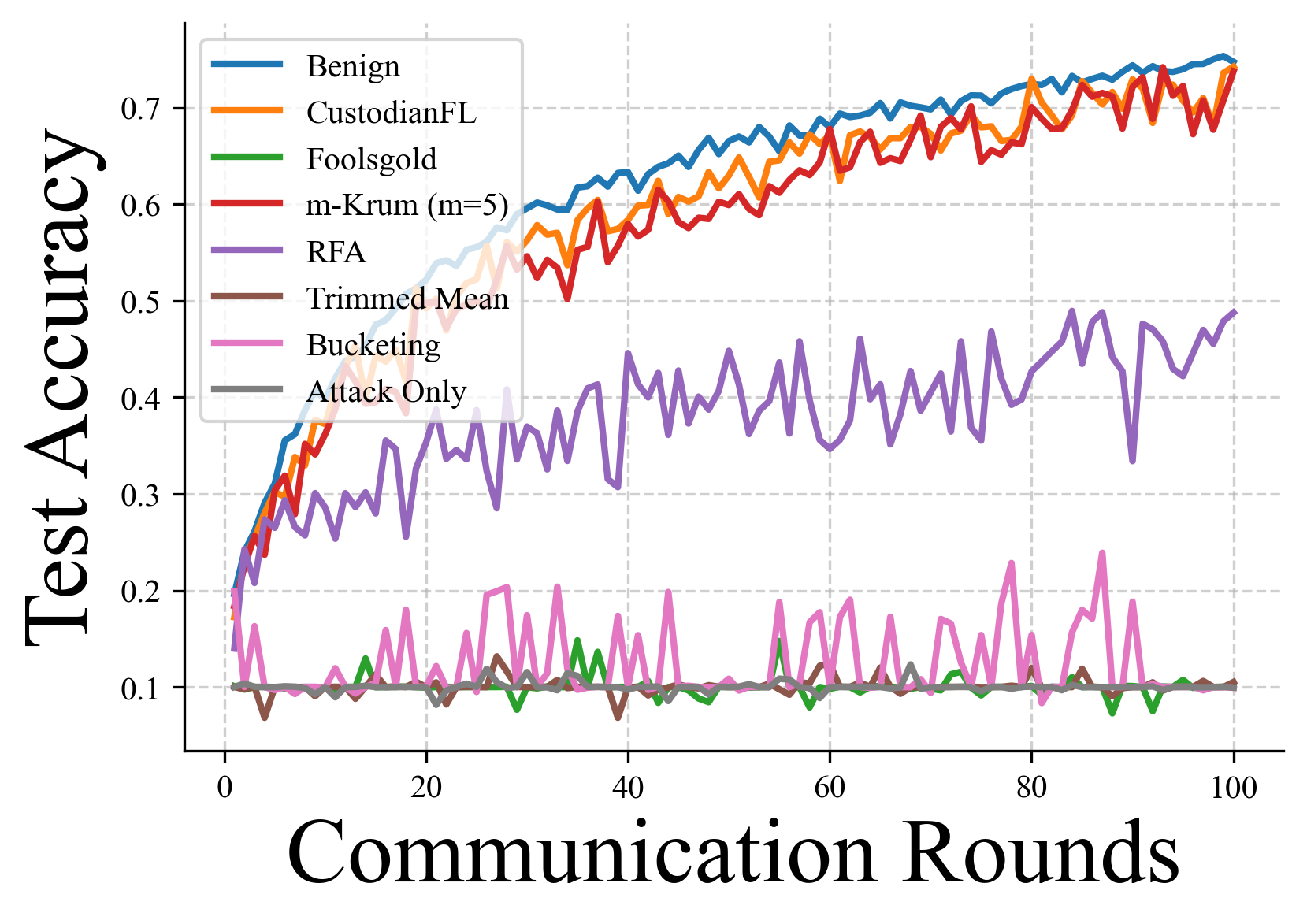}
            \caption{10 clients.}
            \label{fig:resnet20_adaptive_exp_4adv}
        \end{subfigure}\hfill
        \begin{subfigure}[t]{0.496\textwidth}
            \centering
            \includegraphics[width=\linewidth]{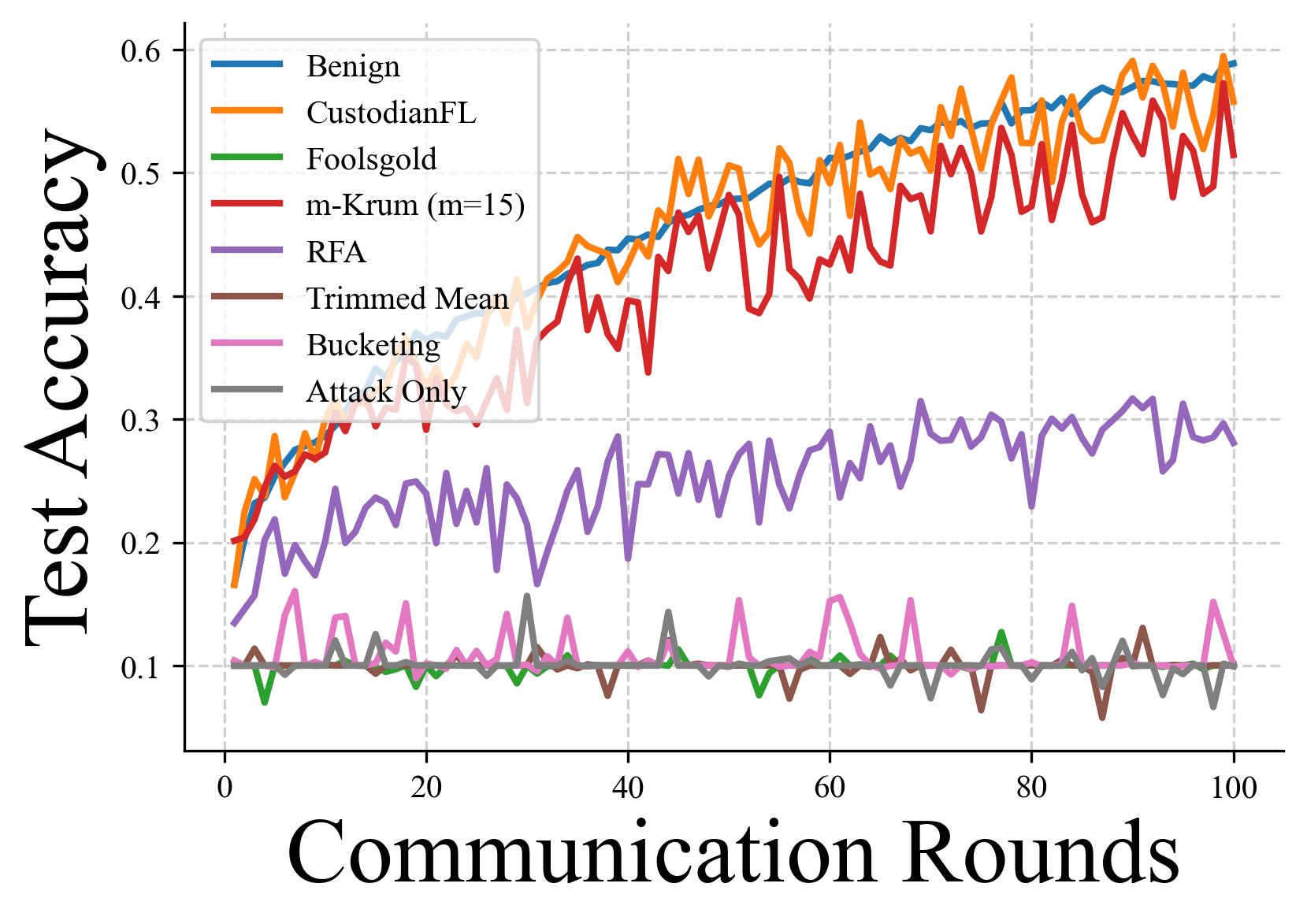}
            \caption{30 clients.}
            \label{fig:resnet20_adaptive_exp_30clients}
        \end{subfigure}

        \caption{Performance under different \# of clients.}
        \label{fig:adaptive_clients}
    \end{minipage}\hfill
    \begin{minipage}[t]{0.496\textwidth}
        \centering
        \begin{subfigure}[t]{0.496\textwidth}
            \centering
            \includegraphics[width=\linewidth]{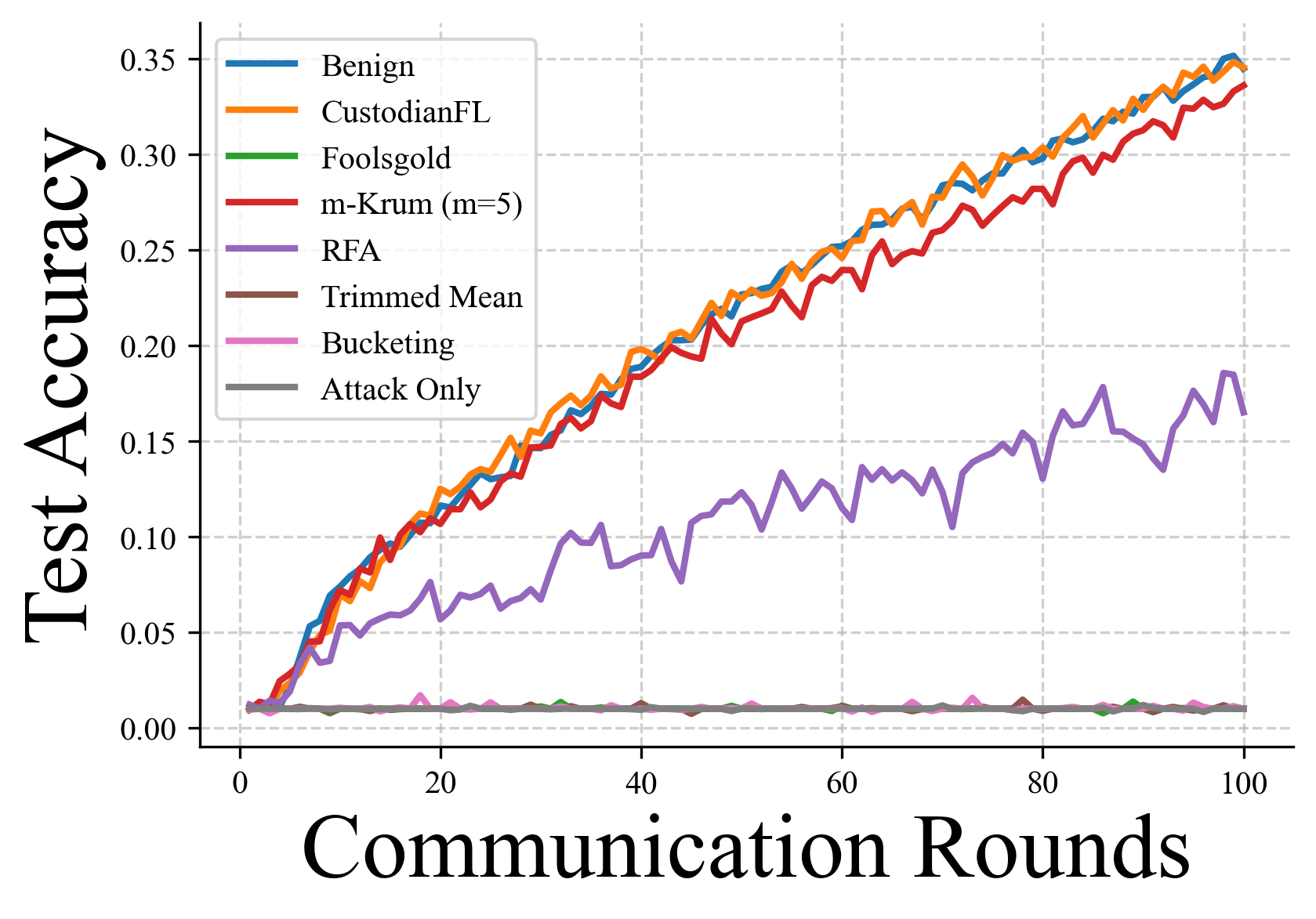}
            \caption{ResNet50, CIFAR100.}
            \label{fig:resnet56_adaptive_exp_4adv} 
        \end{subfigure}\hfill
        \begin{subfigure}[t]{0.496\textwidth}
            \centering
            \includegraphics[width=\linewidth]{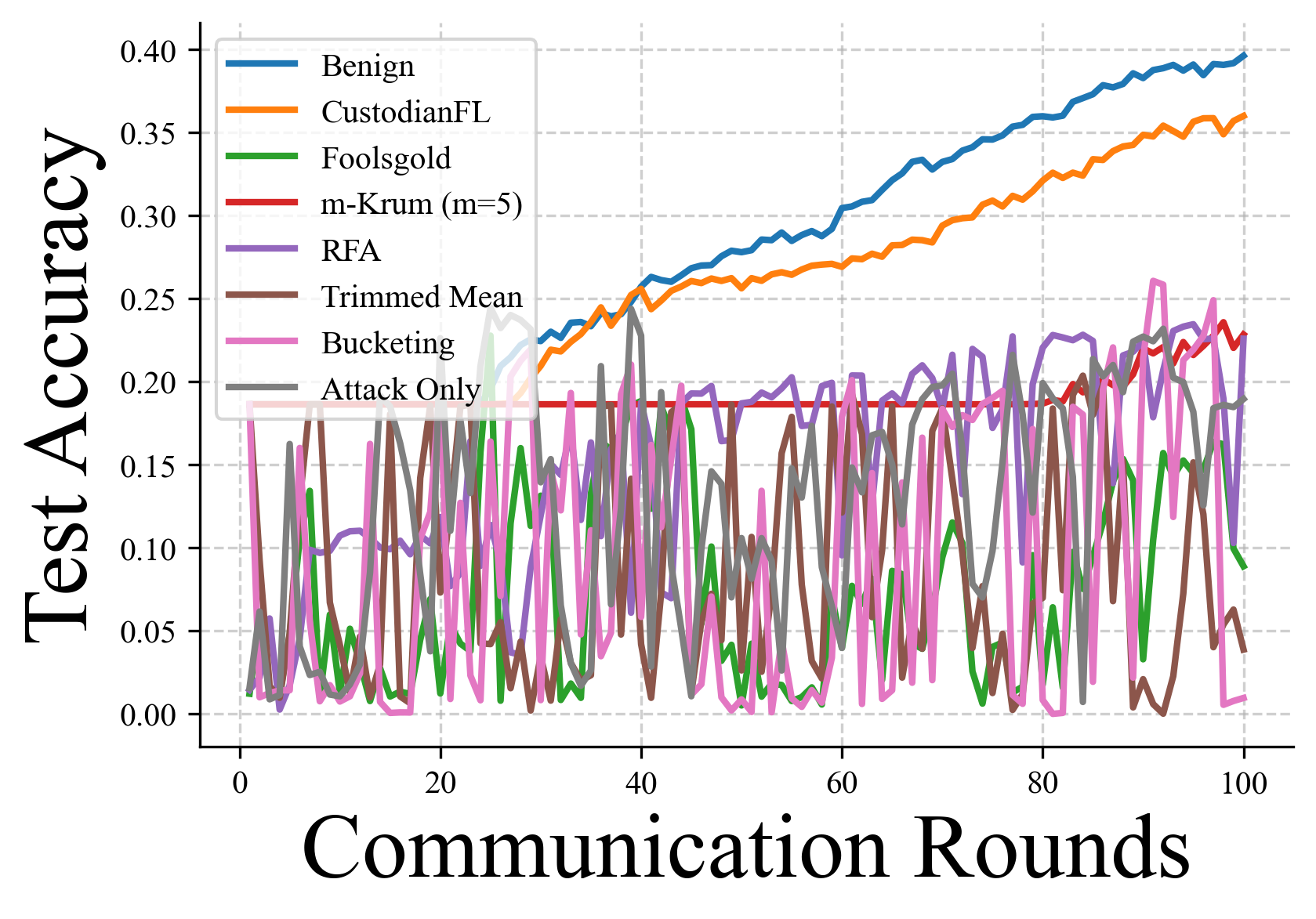}
            \caption{RNN, Shakespeare.}
            \label{fig:nlp_adaptive_exp_4adv}
        \end{subfigure}

        \caption{Performance under different tasks.}
        \label{fig:adaptive_tasks}
    \end{minipage}
\end{figure*}

\noindent\textbf{Exp 4: Varying the percentage of malicious clients.}
This experiment evaluates the impact of varying number of malicious clients on test accuracy. 
We use random Byzantine attack and set the percentage of malicious clients to 20\% and 40\%. 
We also include a baseline case where all clients are benign. As shown in~\cref{fig:malicious_num_exp}, 
the test accuracy remains relatively consistent across different cases, as in each FL training round, our approach filters out the local models that tend to be malicious to minimize the negative impacts of malicious client models on aggregation.

{\noindent\textbf{Exp 5: Varying the number of clients. }We explore the impact of the number of clients under the random Byzantine attack. We set the number of clients to 10, 40, 70, and 100, and set the percentage of malicious clients to 40\%. The results, as described in~\cref{fig:vary_client_num}, indicate that in all cases, our approach has high utility and can filter out malicious clients with high accuracy.}

\noindent\textbf{Exp 6: Evaluations on Byzantine attacks. }We compare our approach with the state-of-the-art defenses and set 10\% of the clients as malicious. We include a ``benign'' case with no activated attack or defense as a baseline.
The results for the random weight Byzantine attack (\cref{fig:compare_defenses_random_byzantine}) and the zero weight Byzantine attack (\cref{fig:byzantine_zero})
demonstrate that our approach (shown in orange color) effectively mitigates the negative impact of the attacks and significantly outperforms the other defenses, by achieving a test accuracy much closer to the benign case.

\noindent\textbf{Exp 7: Evaluations on backdoor attacks. }We compare our approach with the state-of-the-art defenses  and set 10\% of the clients as malicious. Considering that the label flipping attack is subtle as it manipulates local training data and produces malicious local models that are challenging to detect, we set the parameter $\lambda$ to 2 to produce a tighter boundary.
The results for the label flipping attack and model replacement backdoor attack are shown in~\cref{fig:label_flipping_exp} and~\cref{fig:model_replacement}, respectively. Results show that our approach is effective against backdoor attacks, with the test accuracy much closer to the benign case compared to the baseline defenses.

\noindent\textbf{Exp 8: Evaluations on different attack frequencies. }We configure attacks to occur only during specific rounds to evaluate the effectiveness of the proposed two-phase approach. The total number of attack rounds is set to 10, 40, 70, and 100, respectively. We then fix the number of attack rounds to 40 and compare our approach with the state-of-the-art defenses. The results in~\cref{fig:selected_attack} and~\cref{fig:40attack_rounds} show that our method effectively mitigates the impact of the adversarial attacks, ensuring minimal accuracy loss and robust performance even under different attack rounds.

\noindent\textbf{Exp 9: Evaluations on different tasks. }We evaluate the defenses against the random mode of the Byzantine attack with different models and datasets, including: \textit{i}) ResNet-20 + Cifar10, \textit{ii}) ResNet-56 + Cifar100,  and \textit{iii}) RNN + Shakespeare. 
The results in~\cref{fig:resnet20_exp},~\cref{fig:resnet56_exp}, and~\cref{fig:nlp_exp} in Appendix show that 
our approach outperforms the baseline defenses by effectively filtering out poisoned local models, with a test accuracy close to the benign scenarios.
Moreover, some defenses may fail in some tasks, \textit{e}.\textit{g}., $m$-Krum fails in RNN in~\cref{fig:nlp_exp}, as those methods either select a fixed number of local models or re-weight the local models in aggregation, 
which potentially eliminates some local models that are important to the aggregation, leading to an unchanged test accuracy in later FL rounds.

\noindent\textbf{Exp 10: Evaluations against adaptive attacks with different number of clients. }
We evaluate our approach with 10 and 30 clients and compare it to the other defenses, as shown in~\cref{fig:resnet20_adaptive_exp_4adv} and~\cref{fig:resnet20_adaptive_exp_30clients}. In both scenarios, our method achieves high test accuracy that is close to the benign case and consistently outperforms other approaches, demonstrating its strong robustness against adaptive attacks regardless of the number of clients.

\noindent\textbf{Exp 11: Evaluations against adaptive attacks across different tasks. }
We compare our approach with other defenses on the following tasks: \textit{i}) ResNet50 with CIFAR100; and \textit{ii}) RNN with the Shakespeare dataset; See \cref{fig:resnet56_adaptive_exp_4adv} and~\cref{fig:nlp_adaptive_exp_4adv}.
The results show that our approach consistently outperforms other defenses across different tasks, achieving test accuracy close to the benign case. This highlights the effectiveness and generalizability of our method across different tasks.

\begin{figure*}[htbp]
\begin{subfigure}{0.32\textwidth}
        \centering
        \includegraphics[width=\textwidth]{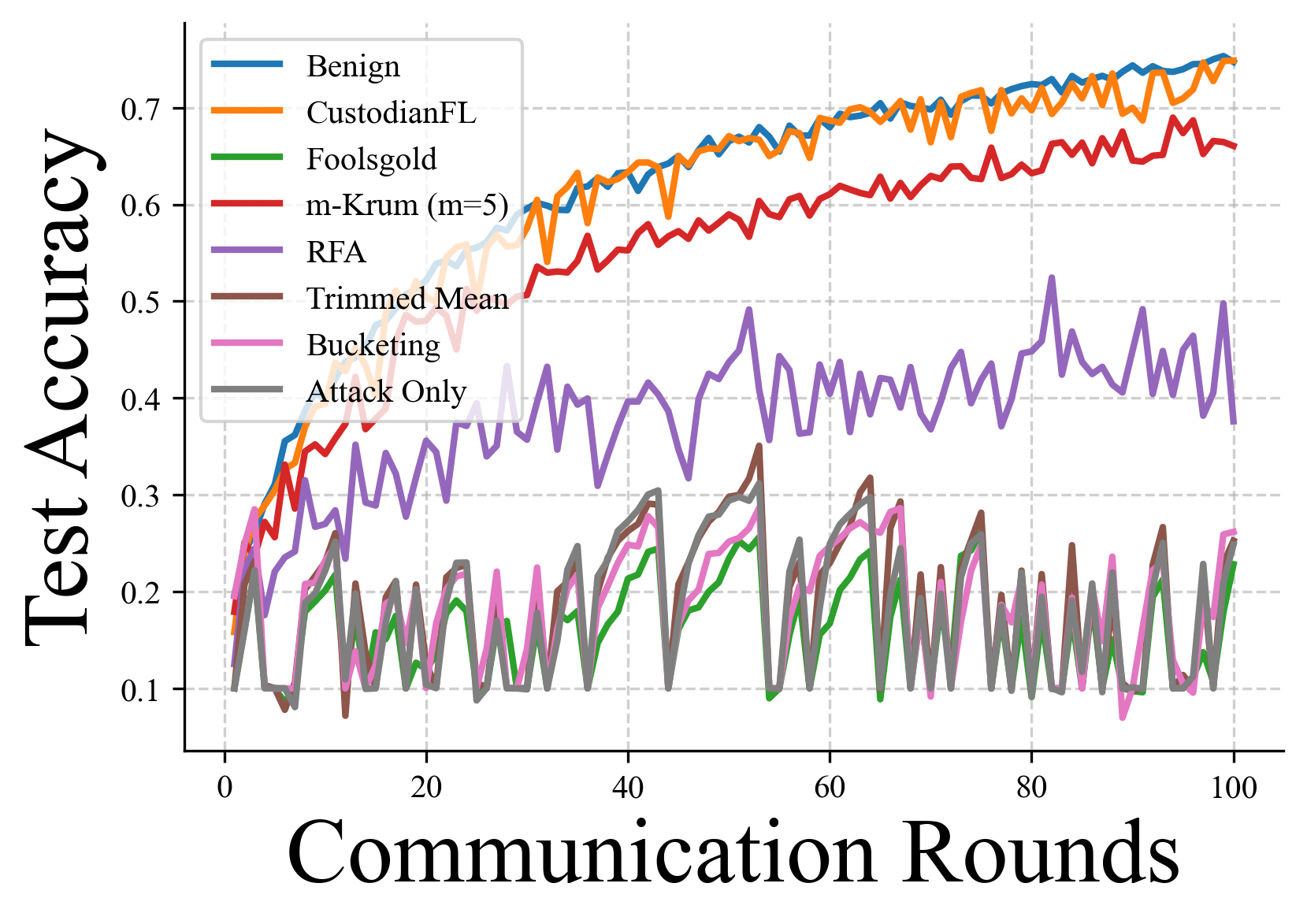}
        \caption{CV: ResNet20 + CIFAR10. }\label{fig:resnet20_adaptive_exp_40random_attacks_4adv}
    \end{subfigure}%
    \hfill
    \centering
    \begin{subfigure}{0.32\textwidth}
        \centering
        \includegraphics[width=\textwidth]{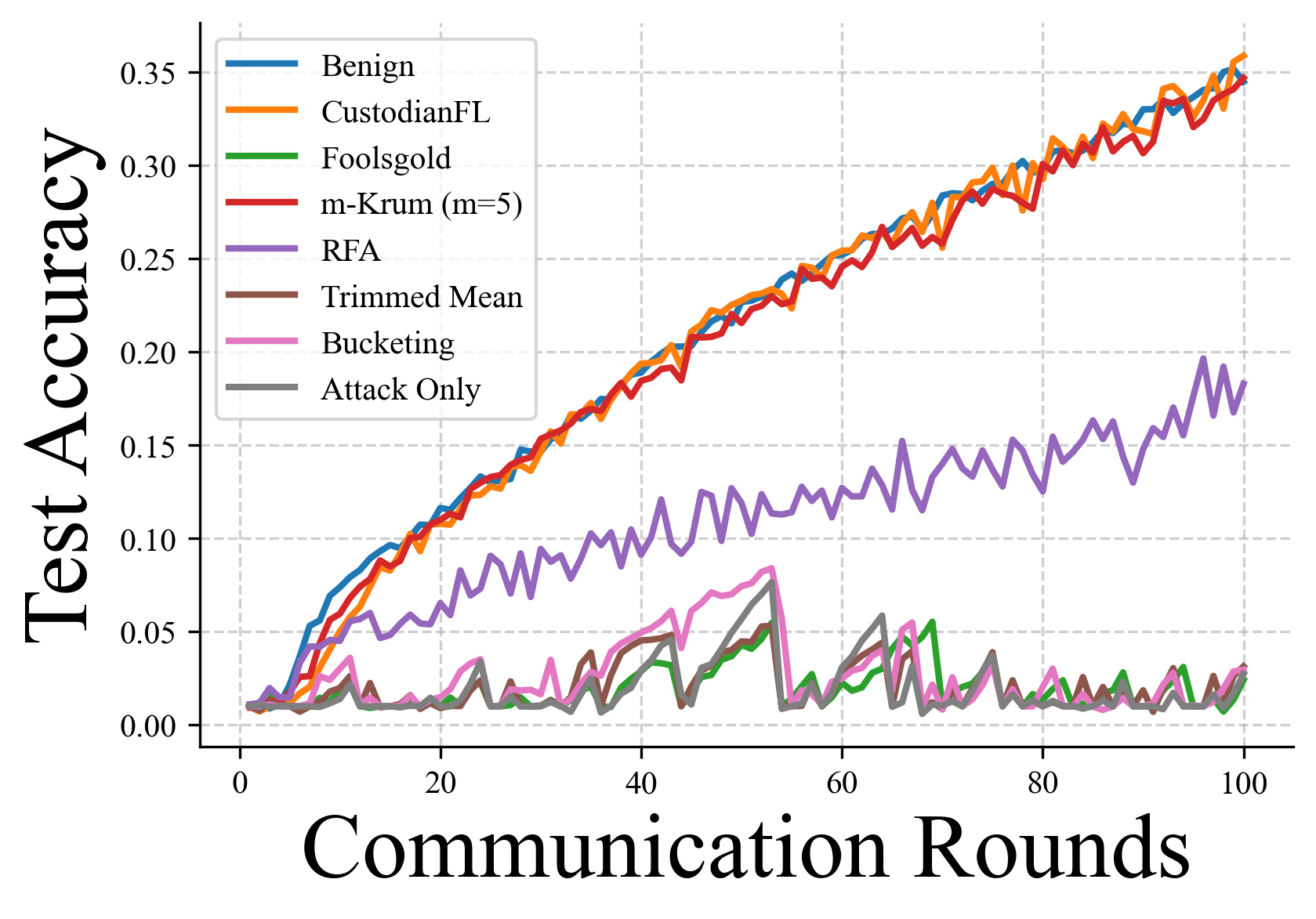}
        \caption{CV: ResNet50 + CIFAR100. }\label{fig:resnet56_adaptive_exp_40random_attacks_4adv}
    \end{subfigure}%
    \hfill
    \begin{subfigure}{0.32\textwidth}
        \centering
        \includegraphics[width=\textwidth]{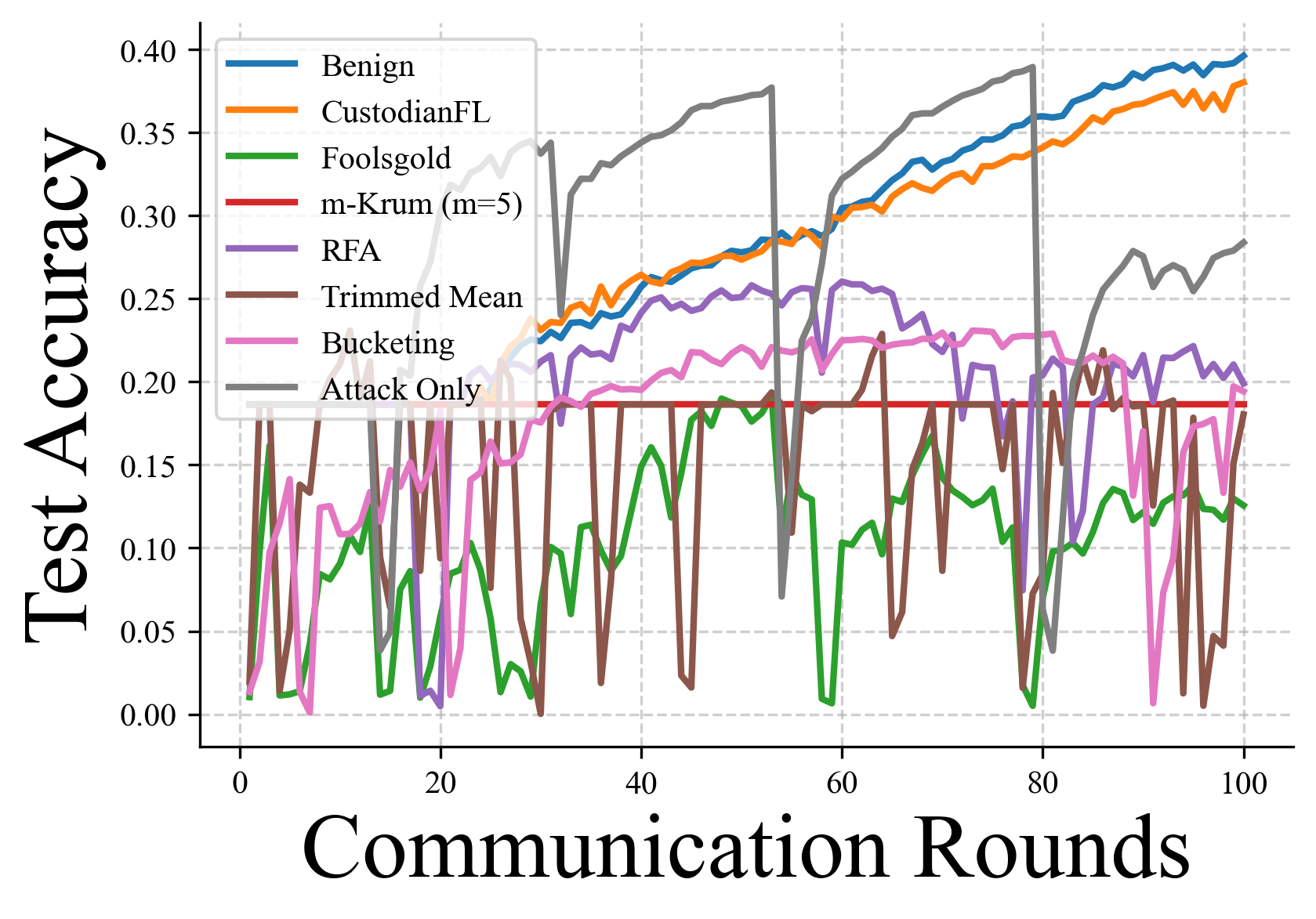}
        \caption{NLP: RNN + Shakespeare. }\label{fig:nlp_adaptive_exp_40random_attacks_4adv}
    \end{subfigure}%
    \caption{Evaluations under a different attack frequency across different tasks.}
\end{figure*}

\begin{table*}[h]
\caption{Cost of ZKP of different models.}
\label{tab: zkp_anomaly_detect}
\centering
\begin{threeparttable}
\begin{tabular}{l|c|c|c|c}
\toprule
\textbf{Model} & Stage 1 Circuit Size & Stage 2 Circuit Size & Proving Time (s) & Verification Time (ms) \\
\midrule
CNN & 476,160 & 795,941 & 33 (12 + 21) & 3 \\
RNN & 1,382,400 & 2,306,341 & 96 (34 + 62) & 3 \\
ResNet-56 & 1,536,000 & 2,562,340 & 100 (37 + 63) & 3 \\
\bottomrule
\end{tabular}
\begin{tablenotes}
\item Bracketed times denote duration for cross-round detection and cross-client detection. 
\end{tablenotes}
\end{threeparttable}
\end{table*}

\noindent\textbf{Exp 12: Evaluations against adaptive attacks under a different attack frequency. }
We set the adaptive attack to occur randomly in 40 out of 100 FL rounds. We
compare our approach with other defenses on different tasks, including: \textit{i}) a CV task: ResNet20 with CIFAR10; \textit{ii}) a CV task: ResNet50 with CIFAR100; and \textit{iii}) an NLP task: RNN with the Shakespeare dataset. The results are demonstrated in \cref{fig:resnet20_adaptive_exp_40random_attacks_4adv},~\cref{fig:resnet56_adaptive_exp_40random_attacks_4adv}, and~\cref{fig:nlp_adaptive_exp_40random_attacks_4adv}. The results show that the performance of \name maintains high performance even when attacks occur randomly, indicating the effectiveness of our method in accurately identifying and removing malicious local models under varying attack frequencies.

\noindent\textbf{Exp 13: Evaluations of ZKP verification.}
We implement a prover's module which contains JavaScript code to generate witness for the ZKP, as well as to perform fixed-point quantization. 
We include CNN, RNN, and ResNet-56 in our evaluations. 
Specifically, we only pull out parameters of the importance layer to represent the whole model to reduce complexity. For instance, the second-to-last layer of the CNN model contains only $7,936$ trainable parameters, as opposed to $1,199,882$ should we use the entire model. 
We report the results in Table~\ref{tab: zkp_anomaly_detect}. 
The results show that the proving is efficient as we utilize importance layers, instead of entire models, for computation. 

\section{Related Works}

Robust learning and the mitigation of adversarial behaviors in FL has been extensively explored~\cite{krum,yang2019byzantine, foolsgold, rfa, he2022byzantine,karimireddy2020byzantine,sun2019can,fu2019attack,ozdayi2021defending,sun2021fl,yin2018byzantine,chen2017distributed,bulyan_guerraoui2018hidden,xie2020slsgd,li2020learning,cao2020fltrust}. 
Some approaches keep several local models that are more likely to be benign in each FL round, \textit{e}.\textit{g}.,~\cite{krum,bulyan_guerraoui2018hidden,yin2018byzantine}, and~\cite{xie2020slsgd}, instead of aggregating all client submissions.
Such approaches are effective, but they keep fewer local models than the real number of benign local models to ensure that all malicious local models are filtered out, causing misrepresentation of some benign local models in the aggregation. This completely wastes the computation resources of the benign clients that are incorrectly removed and thus, changes aggregation results. 
Some approaches re-weight or modify local models to mitigate the impacts of potential malicious submissions~\cite{foolsgold,karimireddy2020byzantine,sun2019can,fu2019attack,ozdayi2021defending,sun2021fl}, 
while other approaches alter the aggregation function or directly modify the aggregation results~\cite{rfa,karimireddy2020byzantine,yin2018byzantine,chen2017distributed}. 
Some approaches detect presence of attacks~\cite{zhang2022fldetector} but requires a number of pre-training rounds and relies heavily on historical client models of previous rounds, making it ineffective when there is limited information on past client models. Moreover, the effectiveness of such approach on early rounds of FL training is challenging, as it might require to set several starting round before detection~\cite{fldetector_implementation}. However, in practice, attacks might happen in early stages of FL training as well.
While these defense mechanisms 
might require unrealistic assumptions or degrade the quality of outcomes due to modifying FL aggregation even in benign cases, thus are not suitable in practical scenarios.

\section{Conclusion}\label{sec:conclusion}

This paper introduces \name, a verifiable anomaly detection method specifically designed for FL systems. Our method introduces an early cross-round detection step that conditionally activates further 
anomaly analysis only when attacks are suspected, thereby minimizing unnecessary interference with benign training. 
\name enhances the reliability of FL systems, fostering trust among FL participants while promoting positive societal impact. 
However, it has certain limitations, \textit{e}.\textit{g}., it does not support asynchronous FL or vertical FL, and the proof generation time for ZKPs remains a bottleneck for wider deployment. Future advancements in ZKP optimization and hardware acceleration are expected to address this issue.

\bibliographystyle{IEEEtranS}
\bibliography{reference}

\newpage
\appendix

\begin{figure}[htbp]
    \centering
    \begin{subfigure}[t]{0.48\columnwidth}
        \centering
        \includegraphics[width=\linewidth]{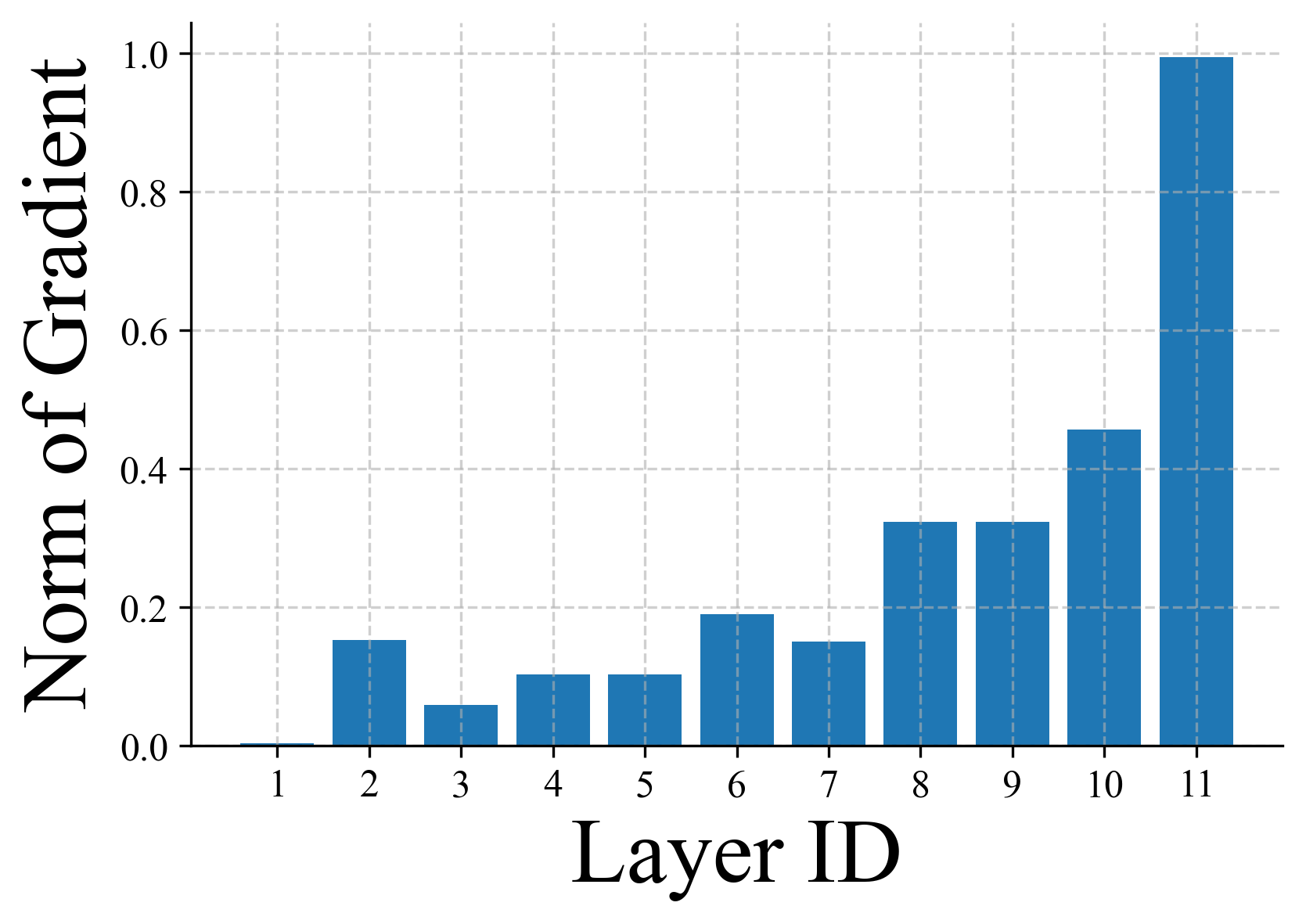}
        \caption{RNN layer sensitivity.}
        \label{fig:rnn_importance_feature}
    \end{subfigure}\hfill
    \begin{subfigure}[t]{0.48\columnwidth}
        \centering
        \includegraphics[width=\linewidth]{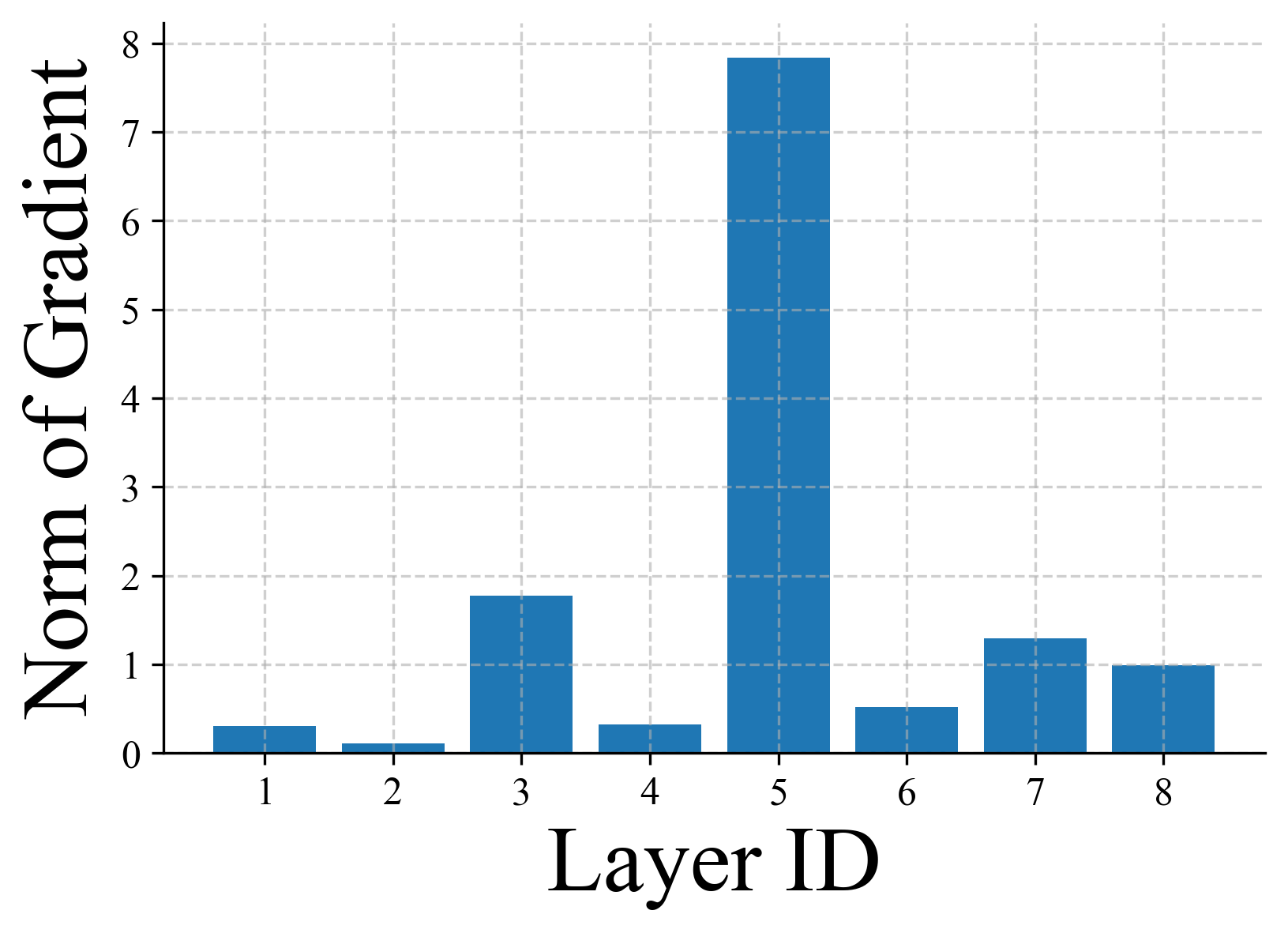}
        \caption{CNN sensitivity.}
        \label{fig:cnn_importance_feature}
    \end{subfigure}
    \caption{Importance layer sensitivity (1/2).}
\end{figure}

\begin{figure}[htbp]
    \centering
    \begin{subfigure}[t]{0.48\columnwidth}
        \centering
        \includegraphics[width=\linewidth]{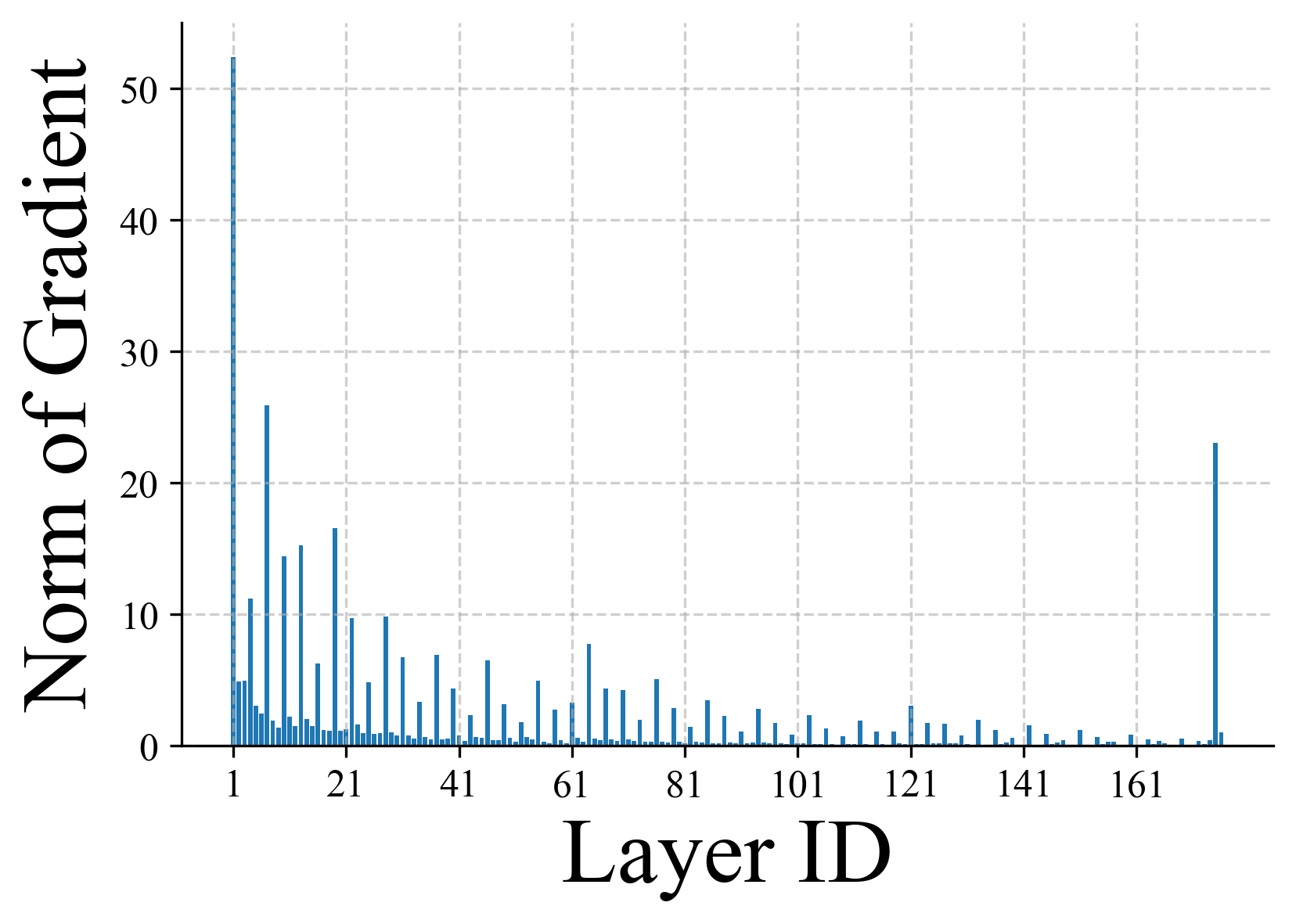}
        \caption{ResNet-56 sensitivity.}
        \label{fig:resnet56_importance_feature}
    \end{subfigure}\hfill
    \begin{subfigure}[t]{0.48\columnwidth}
        \centering
        \includegraphics[width=\linewidth]{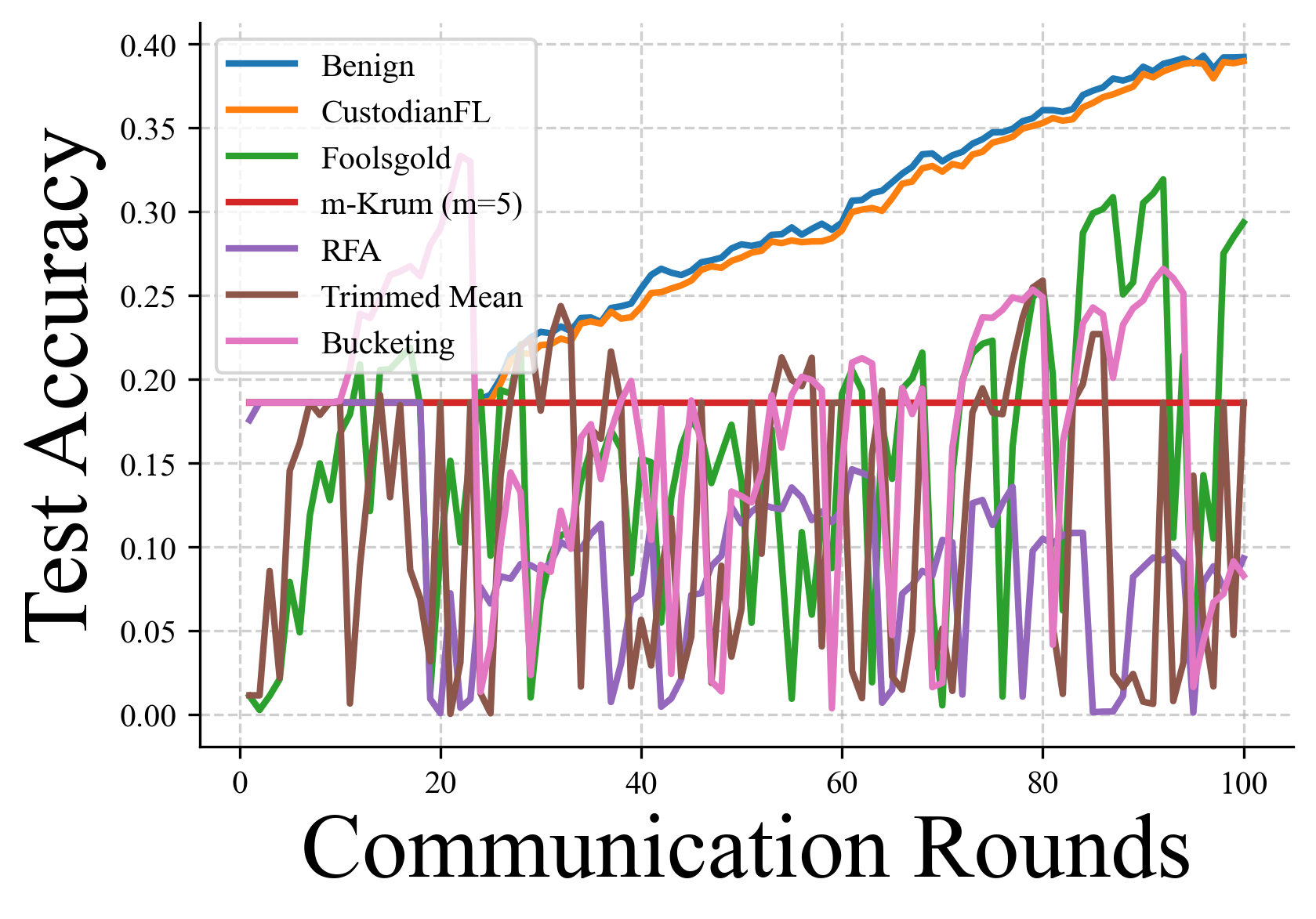}
        \caption{RNN \& Shakespeare.}
        \label{fig:nlp_exp}
    \end{subfigure}
    \caption{Importance layer sensitivity (2/2) and supplementary result.}
    \label{fig:supplementary_results}
\end{figure}

\noindent\textbf{Supplementary Experimental Results. }
The results for the importance layers of RNN, CNN, and ResNet-56 are given in~\cref{fig:rnn_importance_feature},~\cref{fig:cnn_importance_feature}, and~\cref{fig:resnet56_importance_feature}, respectively.
The results for evaluations on RNN and the Shakespeare dataset is shown in \cref{fig:nlp_exp}.


\end{document}